\documentclass[journal, onecolumn, 12pt]{IEEEtran}

\usepackage{graphics}
\usepackage{rotating}
\usepackage{amsfonts}
\usepackage{amsmath}
\usepackage{subfigure}
\usepackage{verbatim}
\usepackage{enumerate}
\usepackage{setspace}
\usepackage{epsfig}
\usepackage{algorithm}
\usepackage{algorithmic}
\usepackage{multirow}
\usepackage{cite}

\graphicspath{{figures/}} \DeclareGraphicsExtensions{.eps,.ps}
\renewcommand{\thefootnote}{\fnsymbol{footnote}}

\doublespace \centerfigcaptionstrue

\begin{document}

\title{Fair and Efficient TCP Access in the IEEE 802.11 Infrastructure Basic Service
Set \footnotemark{$^{\dag}$}}

\author{\singlespace \normalsize \authorblockN{Feyza Keceli, Inanc Inan, and Ender Ayanoglu}\\
\authorblockA{Center for Pervasive Communications and Computing \\
Department of Electrical Engineering and Computer Science\\
The Henry Samueli School of Engineering\\
University of California, Irvine, 92697-2625\\
%Phone: +1 949 824 9797  Fax : +1 949 824 2321\\
Email: \{fkeceli, iinan, ayanoglu\}@uci.edu}}

\maketitle

\footnotetext{$^{\dag}$ This work is supported by the Center for
Pervasive Communications and Computing, and by National Science
Foundation under Grant No. 0434928. Any opinions, findings, and
conclusions or recommendations expressed in this material are
those of authors and do not necessarily reflect the view of the
National Science Foundation.}

\renewcommand{\thefootnote}{\arabic{footnote}}

\begin{abstract}
When the stations in an IEEE 802.11 infrastructure Basic Service
Set (BSS) employ Transmission Control Protocol (TCP) in the
transport layer, this exacerbates per-flow unfair access which is
a direct result of uplink/downlink bandwidth asymmetry in the BSS.
We propose a novel and simple analytical model to approximately
calculate the per-flow TCP congestion window limit that provides
fair and efficient TCP access in a heterogeneous wired-wireless
scenario. The proposed analysis is unique in that it considers the
effects of varying number of uplink and downlink TCP flows,
differing Round Trip Times (RTTs) among TCP connections, and the
use of delayed TCP Acknowledgment (ACK) mechanism. Motivated by
the findings of this analysis, we design a link layer access
control block to be employed only at the Access Point (AP) in
order to resolve the unfair access problem. The novel and simple
idea of the proposed link layer access control block is employing
a congestion control and filtering algorithm on TCP ACK packets of
uplink flows, thereby prioritizing the access of TCP data packets
of downlink flows at the AP. Via simulations, we show that short-
and long-term fair access can be provisioned with the introduction
of the proposed link layer access control block to the protocol
stack of the AP while improving channel utilization and access
delay.
\end{abstract}

\section{Introduction}\label{sec:introduction}

In the IEEE 802.11 Wireless Local Area Networks (WLANs), the
Medium Access Control (MAC) layer employs the Distributed
Coordination Function (DCF) which is a contention-based channel
access scheme \cite{802.11}. The DCF adopts a Carrier Sense
Multiple Access with Collision Avoidance (CSMA/CA) scheme using
binary exponential backoff procedure. In DCF, the wireless
stations, using all equal contention parameters, have equal
opportunity to access the channel. Over a sufficiently long
interval, this results in station-based fair access which can also
be referred as MAC layer fair access. On the other hand,
per-station MAC layer fair access does not simply translate into
achieving per-flow transport layer fair access in the commonly
deployed infrastructure Basic Service Set (BSS), where an Access
Point (AP) serves as a gateway between the wired and wireless
domains. Since the AP has the same access priority with the
wireless stations, an approximately equal bandwidth that an uplink
802.11 station may get is shared among all downlink traffic. This
results in a considerable asymmetry between per-flow uplink and
downlink bandwidth.

The network traffic is currently dominated by data traffic mainly
using Transmission Control Protocol (TCP) in the transport layer.
TCP employs a reliable bi-directional communication scheme. The
TCP receiver returns TCP ACK packets to the TCP transmitter in
order to confirm the successful reception of the data packets. In
the case of multiple uplink and downlink flows in the WLAN,
returning TCP ACK packets of upstream TCP data are queued at the
AP together with the downstream TCP data packets. When the
bandwidth asymmetry in the forward and reverse path builds up the
queue in the AP, the dropped packets impair the TCP flow and
congestion control mechanisms which assume equal transmission rate
both in the forward and the reverse path \cite{Balakrishnan99}.
%When the effects of the bandwidth asymmetry in the infrastructure
%BSS couples with the bi-directional and reliable data delivery
%mechanisms of TCP, TCP connections competing for the shared media
%of the WLAN experience alleviated unfair resource allocation where
%the downlink TCP connections starve in terms of throughput.
As will be described in more detail in Section
\ref{subsec:problem}, unfair bandwidth allocation is observed
between not only uplink and downlink TCP flows but also individual
uplink TCP flows.

A solution for resolving the unfair access problem in the 802.11
BSS for TCP is limiting the TCP packet source rate for all flows
such that no packet drops occur at the AP. This simply translates
into limiting the maximum congestion window size of each TCP
congestion. In this paper, we propose a simple analytical method
to calculate the TCP congestion window limit that prevents packet
drops from the AP queue. The proposed analysis shows that this
window limit can be approximated by a simple linear function of
the bandwidth of the 802.11 WLAN, the number of unlink and
downlink flows, the wired link delay of the TCP connection, the
MAC buffer size of the AP, and the number of TCP data packets each
TCP ACK packet acknowledges. The proposed analysis is generic so
that it considers varying number of uplink and downlink TCP flows,
the use of delayed TCP ACK algorithm, and varying Round Trip Times
(RTTs) among TCP connections. Via simulations, we show that the
analytically calculated congestion window setting provides fair
access and high channel utilization. As we will also describe, the
proposed analysis framework can also be used for buffer sizing at
the AP in order to provision fair TCP access.

Motivated by the findings of the proposed analysis and pointing
out the potential practical limitations of implementation, we also
design a novel link layer access control block to be employed at
the AP. The control block manages the limited AP bandwidth
intelligently by prioritizing the access of the TCP data packets
of downlink flows over the TCP ACK packets of uplink flows. This
is achieved by employing a congestion control and filtering
algorithm on the TCP ACK packets of uplink flows. The specific
algorithm parameters are quantified based on the measured average
downlink data transmission rate. We test the performance of the
protocol stack enhanced with the proposed access control block in
terms of transport layer fairness and throughput via simulations.
The simulation results show that fairness and high channel
utilization can be maintained in a wide range of scenarios.

The rest of this paper is organized as follows. We illustrate the
TCP unfairness problem and provide a brief literature review on
the subject in Section \ref{sec:background}. Section
\ref{sec:analysis} describes the proposed analytical method to
calculate the TCP congestion window limit that prevents packet
drops from the AP queue and provides per-flow fair access in the
WLAN.
%The performance evaluation of
%the proposed analysis is the topic of Section
%\ref{sec:analysis_evaluation}.
Section \ref{sec:proposed} describes the proposed link layer
access control block which uses ACK congestion control and
filtering for fair access provisioning and evaluates its
performance.
%We evaluate the performance of the proposed
%link layer access scheme in Section \ref{sec:simulations}.
We provide our concluding remarks in Section \ref{sec:conclusion}.

\section{Background} \label{sec:background}

\subsection{TCP Fairness} \label{subsec:AIMDfairness}

The congestion avoidance mechanism adopted in TCP can be
characterized by an Additive Increase Multiplicative Decrease
(AIMD) algorithm \cite{Chiu89}. In the congestion avoidance phase,
the congestion window is increased by one at every RTT and is
decreased by half (multiplicative decrease) when a packet loss is
detected.

Consider a simple scenario where two flows share a single
bottleneck link whose capacity is $C$ and they have the same RTT.
Assume that both flows are operating in congestion avoidance
phase. Let $R_{i}$ denote the throughput of TCP flow$_{i}$,
$i={1,2}$.

\begin{figure}[t]
\centering \includegraphics[width = 1.0\linewidth]{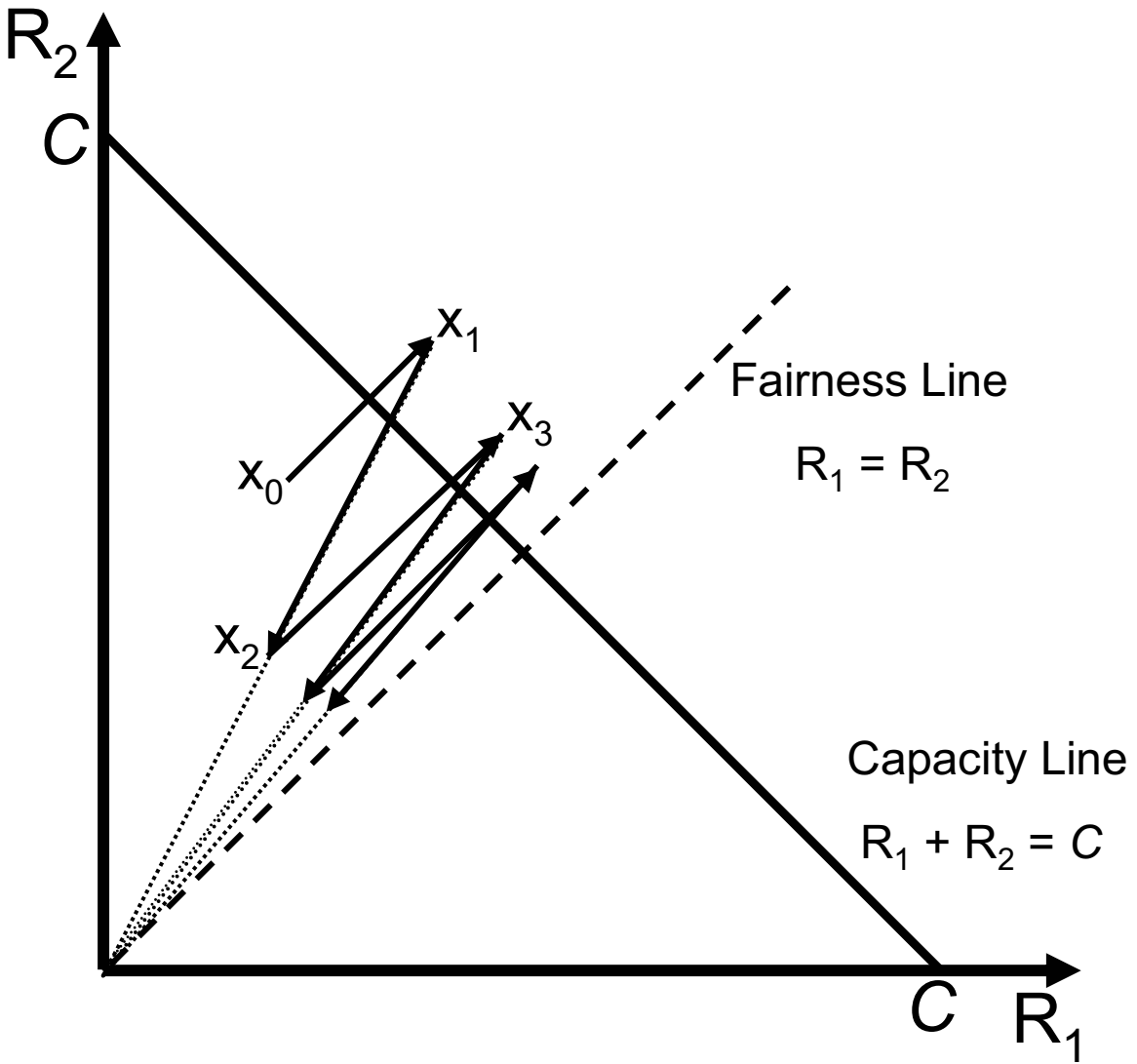}
\caption{AIMD convergence to fair share of the bandwidth}
\label{fig:AIMDfairness}
\end{figure}

In Fig. \ref{fig:AIMDfairness}, x- and y-axis denote the
throughput each flow achieves. Fig. \ref{fig:AIMDfairness} also
shows the system capacity limit and the fair throughput lines.
Initially, suppose that the values of congestion windows for both
flows are such that the throughput pair $(R_{1},R_{2})$ is
achieved as shown by point $x_{0}$ in Fig. \ref{fig:AIMDfairness}.
Note that the process described below is independent of where this
initial point lies. Since $R_{1}+R_{2}<C$, packet losses rarely
occur and both flows increase their congestion windows in an
additive increase manner. This increase corresponds to the line in
Fig. \ref{fig:AIMDfairness} which connects point $x_{0}$ to point
$x_{1}$. As both flows increase their congestion windows in the
same rate, the slope of this line is 1/2. At point $x_{1}$, since
$R'_{1}+R'_{2}>C$, packet losses occur. On detecting the packet
loss, both flows decrease their congestion windows in a
multiplicative manner; to point $x_{2}$, $(R'_{1}/2,R'_{2}/2)$.
This process of alternating increases and decreases continues.
But, eventually, the point $(R_{1}, R_{2})$ reaches the fair
throughput line and stays always on this line. The fluctuation
along this line continues without converging to (0.5C, 0.5C).
Therefore, the AIMD algorithm can provide fair bandwidth sharing
at the expense of oscillations in the throughput.

The AIMD algorithm can be represented as the following generalized
form.
\begin{equation}
\label{eq:AIMD} \setlength{\nulldelimiterspace}{0pt} W[n] =
\left\{ \\
%\begin{IEEEeqnarraybox}[\relax][c]{ll}
\begin{array}{ll}
W[n-1]+\alpha, & ~ {\rm if}~ AI \\
\beta W[n-1], & ~ {\rm if} ~ MD.
\end{array}
%\end{IEEEeqnarraybox}
\right.
\end{equation}
\noindent where $W[N]$ denotes the congestion window value at the
$N_{th}$ transmission round, $\alpha>0$, and $0<\beta<1$. For TCP,
the additive increase and the multiplicative decrease factors,
$\alpha$ and $\beta$ in (\ref{eq:AIMD}), are set to 1 and 0.5,
respectively, as described previously. A higher value of $\alpha$
increases the convergence rate to the fair throughput. Similarly,
a higher value of $\beta$ reduces the oscillations in the
throughput after fair share is achieved.

%The research in [45] investigates the effect of these parameters
%on the effectiveness and responsiveness of the generic AIMD
%algorithm. Similarly, a more generalized AIMD algorithm, termed as
%a binomial congestion control algorithm, has been proposed and its
%fairness has been analyzed in [46].

If our assumption that all the flows have the same RTT does not
hold, the AIMD algorithm cannot assure fairness. A flow with
smaller RTT updates its congestion window quickly and tends to get
more bandwidth compared to a flow with larger RTT. Thus, the TCP
congestion control shows unfairness among flows with different
RTTs.

\subsection{TCP Unfairness in the 802.11
WLAN}\label{subsec:problem}

In the 802.11 WLAN, a bandwidth asymmetry exists between
contending upload and download flows. This is due to the fact that
the MAC layer contention parameters are all equal for the AP and
the stations. If $N$ stations and an AP are always contending for
the access to the wireless channel
(saturation\footnotemark\footnotetext{Saturation is the limit
reached by the system when each station \textit{always} has a
packet to transmit. Conversely, in nonsaturation, the
(nonsaturated) stations experience idle times since they sometimes
have no packet to send.}), each host ends up having approximately
$1/(N +1)$ share of the total transmit opportunities over a long
time interval. This results in $N/(N + 1)$ of the transmissions
being in the uplink, while only $1/(N + 1)$ of the transmissions
belong to the downlink flows.

%The uneven bandwidth share results in downlink flows experiencing
%significantly lower throughput and larger delay. The congestion at
%the AP may result in considerable packet loss depending on the
%size of interface buffers. This is the transport layer
%uplink/downlink unfairness problem in the WLAN.

%The results may even be more catastrophic in a scenario consisting
%of TCP flows.
% TCP flow and congestion control mechanisms run on
%bi-directional communication to ensure reliable transfer of the
%TCP data. The transmission data rate is adjusted according to the
%network capacity.
%The TCP receiver returns TCP ACK packets to the TCP transmitter in
%order to confirm the successful reception of the data packets. In
%the case of multiple uplink and downlink flows in the WLAN,
%returning TCP ACKs of upstream TCP data are queued at the AP
%together with the downstream TCP. When the bandwidth asymmetry in
%the forward and reverse path builds up the queue in the AP, the
%dropped packets impair the TCP flow and congestion control
%mechanisms which assume equal transmission rate both in the
%forward and the reverse path \cite{Balakrishnan99}.

This bandwidth asymmetry in the forward and reverse path may build
up the AP queue resulting in packet drops. As previously stated,
upstream TCP ACKs and downstream TCP data are queued at the AP
together. Any TCP data packet that is dropped from the AP buffer
is retransmitted by the TCP sender following a timeout or the
reception of duplicate ACKs. Conversely, any received TCP ACK can
cumulatively acknowledge all the data packets sent before the data
packet for which the ACK is intended for, i.e., a consequent TCP
ACK can compensate for the loss of the previous TCP ACK. When the
packet loss is severe in the AP buffer, the downstream flows will
experience frequent timeouts thus congestion window size
decreases, resulting in significantly low throughput. On the other
hand, due to the cumulative property of the TCP ACK mechanism,
upstream flows with large congestion windows will not experience
such frequent timeouts. In the latter case, it is a low
probability that many consecutive TCP ACK losses occur for the
same flow. Conversely, the upstream flows with small congestion
windows (fewer packets currently on flight) may also experience
timeouts and decrease their congestion windows even more.
Therefore, a number of upstream flows may starve in terms of
throughput while some other upstream flows enjoy a high
throughput. In summary, the uplink/downlink bandwidth asymmetry
creates a congestion at the AP buffer which results in unfair TCP
access.

In the first set of experiments, we show that the AIMD congestion
avoidance algorithm of TCP leads to fair access when the bandwidth
asymmetry problem between the forward and backward links does not
exist. We consider a scenario consisting of 15 downlink TCP
connections. Each connection is initiated between a separate
wireless station and a separate wired station where an AP is the
gateway between the WLAN and the wired network. Each station runs
a File Transfer Protocol (FTP) session over TCP. Each station uses
802.11g PHY layer with physical layer (PHY) data rate set to 54
Mbps while the wired link data rate is 100 Mbps. The default DCF
MAC parameters are used \cite{802.11}. The AP buffer size is 100
packets. The receiver advertised congestion window limits are set
to 42 packets for each flow. Note that the scale on the buffer
size and TCP congestion window limit is inherited from
\cite{Pilosof03}. Although the practical limits may be larger, the
unfairness problem exists as long as the ratio of the buffer size
to the congestion window limit is not arbitrarily large (which is
not the case in practice). The packet size is 1500 bytes for all
flows.

\begin{figure}[t]
\centering \includegraphics[width =
1.0\linewidth]{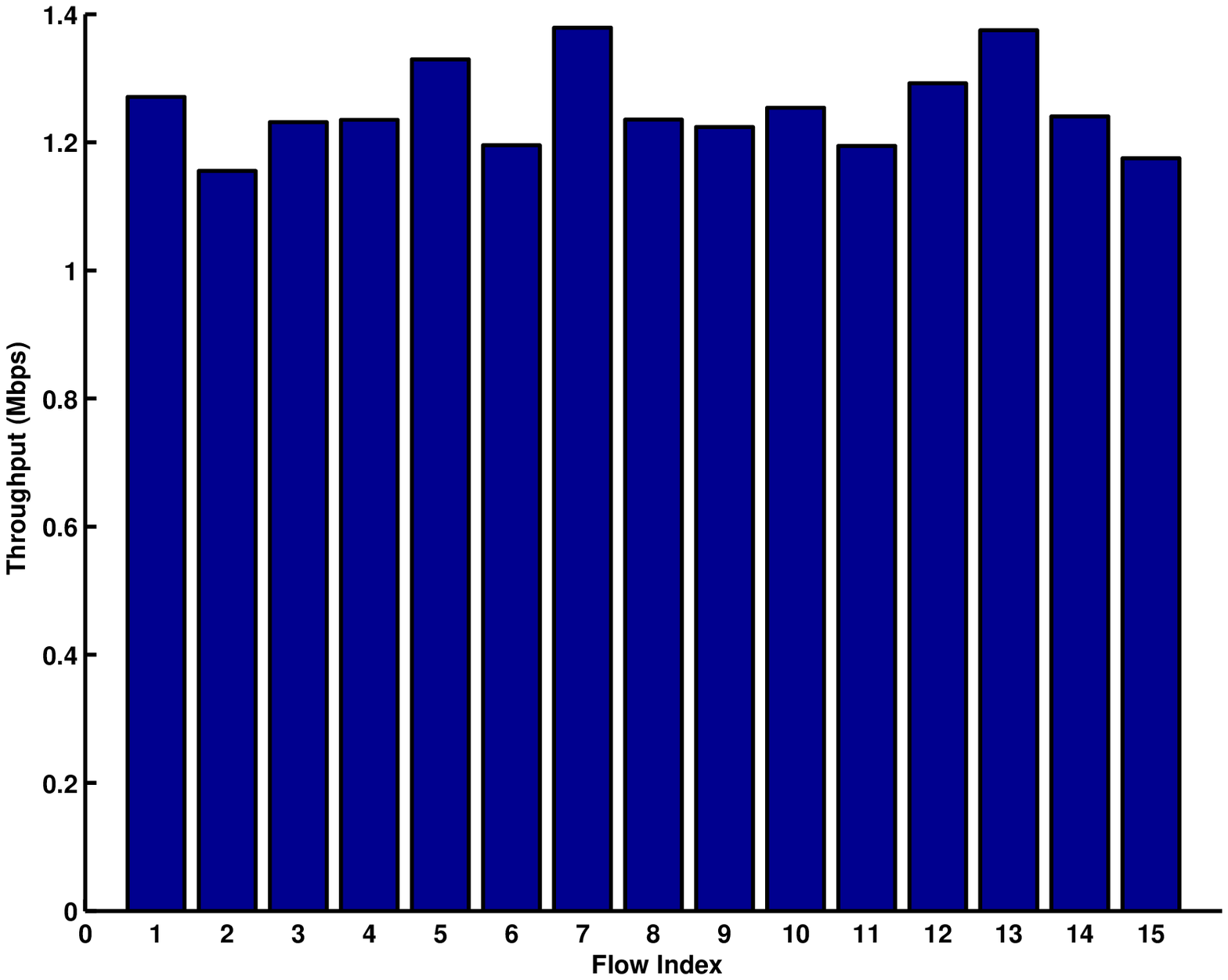} \caption{The
throughput of each TCP connection when there are 15 downlink TCP
connections.} \label{fig:TCPdownlink_fairness}
\end{figure}

Fig. \ref{fig:TCPdownlink_fairness} shows the average throughput
for each downlink TCP connection. The results show that the unfair
access problem does not exist if there are no coexisting uplink
connections (which limit the forward link bandwidth for downlink
connections due to 802.11 uplink/downlink bandwidth asymmetry as
described previously). The results illustrate the fair behavior of
TCP's AIMD congestion avoidance algorithm.

\begin{figure}[t]
\centering \includegraphics[width =
0.8\linewidth]{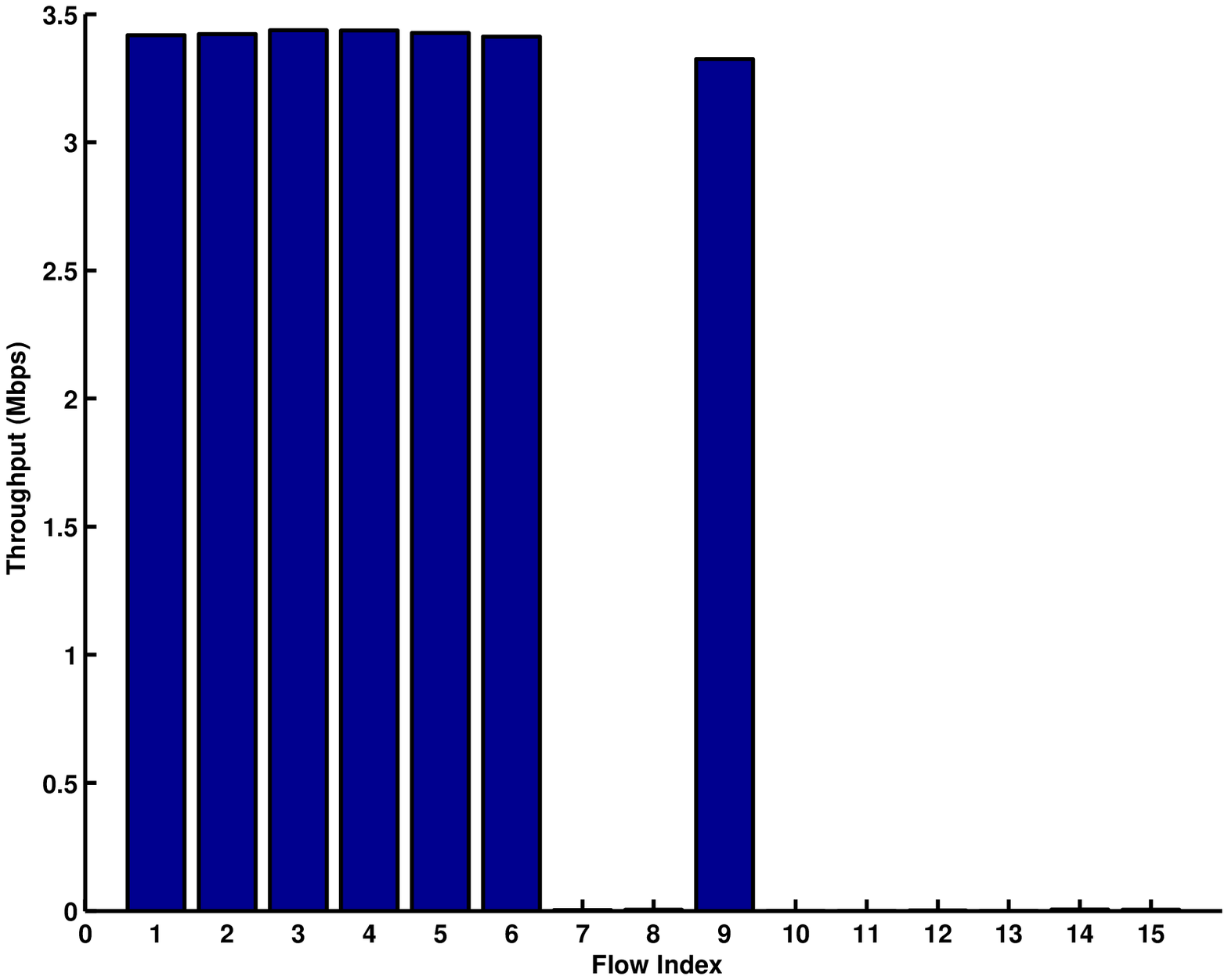} \caption{The
throughput of each TCP connection when there are 15 uplink TCP
connections.} \label{fig:TCPuplink_unfairness}
\end{figure}

When a similar experiment is repeated for the scenario consisting
of only uplink TCP connections, the outcome is very different.
Fig. \ref{fig:TCPuplink_unfairness} shows the average throughput
for each uplink TCP connection a scenario consisting of 15 uplink
TCP connections. The results illustrate the unfairness in the
throughput achieved by the uplink FTP flows when the backward link
bandwidth for TCP ACKs is limited. Eight of the TCP flows starve
in terms of throughput as a result of frequent ACK packet losses
in the backward link at the AP buffer. As described previously in
this section, an ACK packet drop at the AP buffer more likely
results in a congestion window decrease when the flow has a small
congestion window (i.e., a new connection, or a connection
recovering from a recent timeout, etc.). Conversely, a flow with a
higher congestion window size may not be affected because of the
cumulative ACK feature of TCP.

As the comparison of Fig. \ref{fig:TCPdownlink_fairness} and Fig.
\ref{fig:TCPuplink_unfairness} clearly shows, the cumulative
nature of TCP ACKs affects the fair share of the bandwidth
significantly, when the bandwidth asymmetry in between the forward
and backward links exists.

\begin{figure}[t]
\centering \includegraphics[width =
0.8\linewidth]{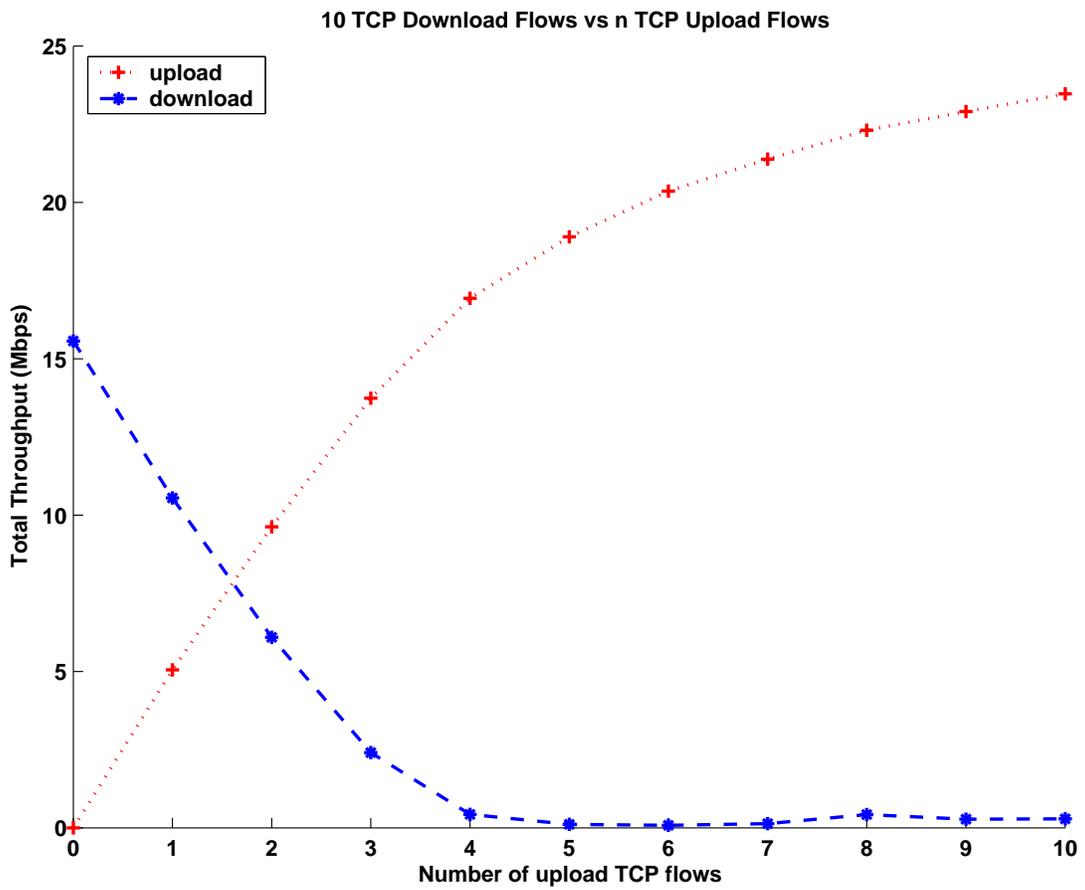} \caption{The total TCP
throughput in the downlink and the uplink when there are 10
download TCP connections and the number of upload TCP connections
varies from 1 to 10.} \label{fig:unfair}
\end{figure}

In the second set of experiments, we show the uplink/downlink
bandwidth asymmetry for DCF and how this is exacerbated if TCP is
employed. We use the same simulation parameters as in the previous
experiment. Fig. \ref{fig:unfair} shows the total TCP throughput
in the downlink and the uplink when there are 10 download TCP
connections and the number of upload TCP connections is varied
from 0 to 10. The unfairness problem between upstream and
downstream TCP flows is evident from the results. For example, in
the case of 2 upload connections, 10 download TCP connections
share a total bandwidth of 6.09 Mbps, while 2 upload TCP
connections enjoy a larger total bandwidth of 9.62 Mbps. As the
number of upload connections increases, the download TCP
connections are almost shut down.

\subsection{Literature Overview}\label{subsec:literature}

The studies in the literature on the unfair access problem in the
802.11 WLAN can mainly be classified into two.

The first group mainly proposes access parameter differentiation
between the AP and the stations to combat the problem. Distributed
algorithms for achieving MAC layer fairness in 802.11 WLANs are
proposed in \cite{Vaidya00}, \cite{Nandagopal00}. Several studies
propose using the traffic category-based MAC prioritization
schemes of IEEE 802.11e standard \cite{802.11e} mainly designed
for Quality-of-Service (QoS) provisioning for uplink/downlink
direction-based differentiation in order to improve fairness and
channel utilization
\cite{Casetti04,Leith05,Freitag06,Tinnirello05_2,Keceli08_ICCEDCA}.
Algorithms that study enhancements on the backoff procedure for
fairness provisioning are proposed in \cite{SWKim05,Jeong05}.
Although MAC parameter differentiation, adaptation, and backoff
procedure enhancements can be effective in fair access
provisioning, the 802.11 hardware (Network Interface Cards (NICs),
APs, etc.) without these capabilities is still widely deployed.
Therefore, in this paper, we focus on techniques that do not
require any changes in the 802.11 standard or in the non-AP
stations and can directly be implemented via simple software
modules in the AP protocol stack.
%A simulation-based analysis is carried out for a specific scenario
%consisting of TCP and audio flows both in the uplink and the
%downlink. An experimental study is carried out in \cite{Leith05}
%to decide on CW and TXOP values of the AP and the stations for a
%scenario with TCP uplink and downlink flows. Both solutions
%propose that individual uplink and downlink streams use separate
%ACs. No guidelines are provided on how to decide on the EDCA
%parameters that achieve fair resource allocation for an arbitrary
%scenario.
%Also, the interaction of TCP flow and congestion control
%mechanisms with the MAC is not addressed.
%In \cite{SWKim05}, it is proposed that the AP accesses the channel
%in Point Interframe Space (PIFS) completion without any backoff
%when the utilization ratio drops below a threshold. Achieving
%weighted fairness between uplink and downlink in DCF is studied
%through mean backoff distribution adjustment in \cite{Jeong05}.

%A mechanism that dynamically tunes CW and TXOP values in order to
%prevent delay asymmetry of realtime UDP flows is proposed in . An
%adaptive priority control mechanism is employed in \cite{Shin06}
%to balance the uplink and downlink delay of VoIP traffic.

The second group focuses on designing higher layer solutions such
as employing queue management, packet filtering schemes, etc.,
especially for TCP. The TCP uplink and downlink asymmetry problem
in the IEEE 802.11 infrastructure BSS is first studied in
\cite{Pilosof03}.
%, where
%the effect of the AP buffer size in the wireless channel bandwidth
%allocation is studied.
The proposed solution of \cite{Pilosof03} is to manipulate
advertised receiver windows of the TCP packets at the AP. In this
paper, we propose a simple analytical model to calculate the
congestion window limit of TCP flows for the generic case of
delayed TCP ACK schemes and varying RTTs among TCP connections.
The results of the proposed analysis can be used in the same way
as proposed in \cite{Pilosof03} for fair and efficient access
provisioning. Per-flow queueing \cite{Wu05} and per-direction
queueing \cite{Ha06} algorithms where distinct queues access the
medium with different probabilities are designed for fair access
provisioning. A rate-limiter block which filters data packets both
in the uplink and the downlink using instantaneous WLAN bandwidth
estimations is proposed in \cite{Melazzi05}. Differing from all of
these techniques, in our previous work, we proposed using
congestion control and filtering techniques on top of the MAC
queue to solve the TCP \textit{uplink} unfairness problem
\cite{Keceli07_ICC}. The work presented in this paper proposes a
novel congestion control and filtering technique which also
considers the TCP downlink traffic. Note that since TCP downlink
traffic load is expected to be larger than the uplink traffic
load, this enhancement is vital for a practical implementation.

An extensive body of work exists relating to the impact of
asymmetric paths on TCP performance \cite{RFC3449,RFC2760} in the
wired link context. The effects of ACK congestion control on the
performance of an asymmetric network are analyzed in
\cite{Balakrishnan97} for wired scenarios consisting of only one
or two simultaneous flows.
%It is noted that any scheme that reduces
%the frequency of ACKs in the wired scheme has potential unwanted
%side effects such as slow congestion window growth (unless byte
%counting is used), increase burstiness on the data side, increase
%the RTT experience by a flow and possibility of negative impact on
%performance during slow start and loss recovery. The proposed
%algorithm in this paper also uses the idea of thinning the ACK
%stream in the constrained channel. We carefully analyze each of
%the stated side effects for the wireless scenario and add the
%combatting capability to the algorithm if necessary.
The effects of forward and backward link bandwidth asymmetry have
been analyzed in \cite{Lakshman97} for a wired scenario consisting
only one flow. Similar effects are also observed in practical
broadband satellite networks \cite{Fairhurst01}. The effects of
delayed acknowledgements and byte counting on TCP performance are
studied in \cite{Allman98}. Several schemes are analyzed in
\cite{Kalampouskas98} for improving the performance of two-way TCP
traffic over asymmetric links where the bandwidths in two
dimensions differ substantially. The ACK compression phenomenon
that occurs due to the dynamics of two-way traffic using the same
buffer is presented in \cite{Zhang91}. In this paper, we design a
novel ACK congestion control and filtering algorithm to be
implemented as a link layer access control block in the protocol
stack at an 802.11 AP. The congestion control and filtering
algorithm is unique in that the parameters of the algorithm are
quantified according to the TCP access characteristics in an
802.11 infrastructure BSS.

%In this paper, we design an ACK congestion control and filtering
%algorithm for the 802.11 infrastructure BSS where an arbitrary
%number of upload and download TCP connections may coexist.

%The functionality is provided by a link layer access control block
%that is introduced right above the MAC layer buffer in the
%protocol stack.

\section{TCP Fairness Analysis} \label{sec:analysis}

%In order to analyze the fairness and the throughput of TCP access
%in the infrastructure BSS, we conduct a simple but comprehensive
%and insightful joint analytical and simulation study.

%In the first part of the analysis, we assume that there are no
%packet losses in any of the TCP connections neither in the forward
%or the backward streams. Moreover, let each TCP connection have
%equal maximum congestion window size. These assumptions are made
%to simplify the analytical approach. In the sequel, we will also
%discuss the cases where these assumptions are released. We will
%discuss the necessary conditions to be satisfied for fair access
%if congestion window sizes of individual TCP connections differ.
%We will show how the packet losses at the AP buffer leads to
%unfair access.

The TCP unfairness problem originating from the uplink/downlink
access asymmetry can be resolved if packet drops at the AP buffer
are prevented such as in the unrealistic case of infinitely long
AP queue. In this case, congestion windows of all TCP flows
whether in the downlink or uplink reach up to the receiver
advertised congestion window limit and stay at this value. This
results in fair access in opposed to the fact that the access is
asymmetric in the 802.11 infrastructure BSS as described in
Section \ref{subsec:problem}. As the infinitely long queue
assumption is unrealistic, the exact same result of no packet
drops can be achieved if TCP senders are throttled by limiting the
number of packets in flight, i.e., the TCP congestion windows are
assigned regarding the available AP bandwidth in the downlink. In
this section, we propose a simple and novel analytical model to
calculate the maximum congestion window limit of each TCP flow
that prevents packet losses at the AP buffer, therefore provides
fair and efficient TCP access in the BSS.

Each random access system exhibits cyclic behavior. The cycle time
is defined as the average duration in which an arbitrary tagged
station successfully transmits one packet on average. Our
analytical method for calculating the TCP congestion window limit
that achieves fair and efficient access is based on the cycle time
analysis previously proposed for 802.11 MAC performance modeling
\cite{Medepalli05},\cite{Inan07_Globecom_Cycle}. The simple cycle
time analysis assesses the asymptotic performance of the DCF
accurately (when each contending AC always has a packet in
service). We use the approach in \cite{Medepalli05} to derive the
explicit mathematical expression for the average DCF cycle time
when necessary. In Section \ref{sec:CTAP}, we will describe the
necessary extensions to employ the cycle time analysis in the
proposed analysis. Due to space limititations, the reader is
referred to \cite{Medepalli05},\cite{Inan07_Globecom_Cycle} for
details on the derivation of cycle time.

We consider a typical network topology where a TCP connection is
initiated between a wireless station and a wired station either in
the downlink or the uplink of the WLAN. The WLAN traffic is
relayed to the wired network through the AP and vice versa. Let
Round Trip Time (RTT) denote the average length of the interval
from the time a TCP data packet is generated until the
corresponding TCP ACK packet arrives. RTT is composed of three
main components as follows.
\begin{itemize}
\item \textbf{Wired Link Delay ($LD$):} The flow-specific average propagation
delay of the packet between the AP and the wired node.
%Since we
%investigate the fairness in the wireless link, we assume that $LD$
%is constant for each connection. In the first part of the
%analysis, we assume each connection has equal $LD$. We will also
%release this assumption later on.
\item \textbf{Queueing Delay
($QD$):} The average delay experienced by a packet at the wireless
station buffer until it reaches to the head of the queue. Note
that due to the unequal traffic load at the AP and the stations,
$QD_{AP}$ and $QD_{STA}$ may highly differ.
\item \textbf{Wireless
Medium Access Delay ($AD$):} The average access delay experienced
by a packet from the time it reaches to the head of the MAC queue
until the transmission is completed successfully.
\end{itemize}
\noindent Then, RTT is calculated as
follows\footnotemark\footnotetext{RTT is calculated as in
(\ref{eq:RTTcalc}) irrespective of the direction of the TCP
connection. On the other hand, specific values of $AD$ and $QD$
depend on the packet size, the number of contending stations, etc.
Therefore, RTT of an uplink connection may differ from RTT of a
downlink connection.}.
\begin{equation}\label{eq:RTTcalc}
RTT = 2 \cdot LD + QD_{AP} + QD_{STA} + AD_{AP} + AD_{STA}
\end{equation}

%The average throughput of TCP connection is determined by the
%number of packets that can be transmitted successfully in one RTT
%(before the TCP ACK of the first packet arrives). Therefore, TCP
%throughput is either limited by the length of RTT or the
%congestion window size.

%As it is discussed in Section \ref{subsec:problem}, at high load,
%the AP can get only $1/N$ share of the total transmissions over a
%long time interval which results in unfair access between the TCP
%connections. On the other hand,

For the first part of the analysis, each TCP data packet is
assumed to be acknowledged by a TCP ACK packet where this
assumption is later released and the delayed TCP ACK algorithm is
considered.

We claim that if the system is to be stabilized at a point such
that no packet drops occur at the AP queue, then the following
conditions should hold.
\begin{itemize}
\item \textit{All the non-AP stations are in nonsaturated
condition.}

Let's assume a station has $X$ packets (TCP data or ACK) in its
queue. A new packet is generated only if the station receives
packets (TCP ACK or data) from the AP (as a result of ACK-oriented
rate control of TCP). Let $Y>1$ users to be active. Every station
(including the AP) sends one packet successfully every cycle time
\cite{Medepalli05}. In the stable case, while the tagged station
sends $Y$ packets every $Y$ cycle time, it receives only one
packet. Note that the AP also sends $Y$ packets during $Y$ cycle
times, but on the average, $Y-1$ of these packets are destined to
the stations other than the tagged one. Therefore, after $Y$ cycle
times, the tagged stations queue size will drop down to $X-Y+1$.
Since $Y>1$, the tagged stations queue will get empty eventually.
A new packet will only be created when the AP sends a TCP packet
to the tagged station which will be served before it receives
another packet (on average). This proves that all the non-AP
stations are in nonsaturated condition if no packet losses occur
at the AP.
\item \textit{The AP contends with at most one station at a time on
average.}

Following the previous claim, a non-AP station (which is
nonsaturated) can have a packet ready for transmission if the AP
has previously sent a packet to the station. There may be
transient cases where the instantaneous number of active stations
may become larger than 1. On the other hand, as we have previously
shown, when $Y>1$, the queue at any non-AP station eventually
empties. If we assume the transient duration being very short, the
number actively contending stations on average is one. Therefore,
at each DCF cycle time, the AP and a distinct station will
transmit a packet successfully.
\end{itemize}

We define $CT_{AP}$ as the duration of the average cycle time
during which the AP sends an arbitrary packet (TCP data or ACK)
successfully. We will derive $CT_{AP}$ in Section \ref{sec:CTAP}.
Let the average duration between two successful packet
transmissions of an arbitrary flow at the AP (or at the non-AP
station) be $CT_{flow}$. Assuming there are $n_{up}$ and
$n_{down}$ upload and download TCP connections respectively, we
make the following approximation based on our claims that the AP
contends with one station on average and the TCP access will be
fair if no packet drops are observed at the AP buffer
\begin{equation}\label{eq:CTflow}
CT_{flow} \cong (n_{up}+n_{down}) \cdot CT_{AP}.
\end{equation}
\noindent As it will be shown by comparing with simulation results
in Section \ref{subsec:analysis_evaluation}, the approximation in
(\ref{eq:CTflow}) leads to analytically correct results.

Then, the throughput of each station (whether it is running an
uplink or a downlink TCP connection) is limited by $1/CT_{flow}$
(in terms of packets per second). We can also write the TCP
throughput using $W_{lim}/RTT$, where we define $W_{lim}$ as the
TCP congestion window limit for a TCP connection.
%We calculate $W_{lim}$ such that if a flow uses a congestion
%window limit larger than $W_{lim}$, this results in packet losses
%at the AP buffer. As described in Section \ref{subsec:problem},
%the drop of a TCP data packet (for downlink TCP connections)
%directly results in a timeout or the generation of duplicate TCP
%ACKs. On the other hand, due to the cumulative nature of TCP ACK
%packets, the uplink flows can still maintain their high throughput
%if the ACK drop ratio does not go over a threshold. As a result,
%unfair resource allocation in the WLAN is observed.
Following our previous claims, $QD_{STA}=0$ (the stations are
nonsaturated), $QD_{AP}=(BS_{AP}-1)\cdot CT_{AP}$ (we consider the
limiting case when the AP buffer is full, but no packet drop is
observed), and $AD_{AP}+AD_{STA}=CT_{AP}$ (the AP contends with
one station on average), where $BS_{AP}$ is the buffer size of the
AP MAC queue. Using $1/CT_{flow}=W_{lim}/RTT$, we find
\begin{equation}\label{eq:w_lim}
W_{lim}= \frac{2\cdot
LD}{CT_{flow}}+\frac{BS_{AP}}{n_{up}+n_{down}}.
\end{equation}
\noindent Note that $CT_{flow}$ is an indication of the bandwidth
at the bottleneck (at the AP). If the data rate exceeds this
bandwidth, the excess data will be queued at the AP, eventually
overflowing the AP buffer. We calculate $W_{lim}$ considering a
full AP buffer, therefore, $W_{lim}$ is the \textit{maximum}
congestion window limit for a TCP connection that prevents the
packet drops at the AP queue of size $BS_{AP}$.

We can make following observations from (\ref{eq:w_lim}).

\begin{itemize}
\item $W_{lim}$ is a function of $LD$. Therefore, $W_{lim}$
is flow-specific and varies among connections with different $LD$.
\item The first term is the effective number of
packets that are in flight in the wired link for any flow, while
the second term is the number of packets that are in the AP buffer
for the same flow.
\end{itemize}

\subsection{Delayed TCP Acknowledgements}

In the delayed TCP ACK mechanism, the TCP receiver acknowledges
every $b$ TCP data packets ($b>1$). A typical value (widely used
in practice) is $b=2$.

The use of delayed TCP ACK mechanism changes the system dynamics.
On the other hand, we still employ our assumption that the AP
contends one station at a time on the average to calculate
$CT_{AP}$. As will be shown by comparison with simulation results
in Section \ref{subsec:analysis_evaluation}, this assumption still
leads to analytically accurate results.

We update (\ref{eq:CTflow}) and (\ref{eq:w_lim}) accordingly for
delayed TCP acknowledgments. Let the average duration between two
successful packet transmissions of the flow at the non-AP station
be $CT_{flow,del}$ when delayed TCP acknowledgment mechanism is
used. Each uplink flow completes the successful transmission of
$b$ packets in an interval of average length $b\cdot
CT_{flow,del}$. When the access is fair, the AP transmits $b$ data
packets for each downlink flow (i.e., a total of $b\cdot
n_{down}$), and one ACK packet for each uplink flow (i.e., a total
of $n_{up}$ ACK packets) during the same interval. Then,

%$CT_{flow,del}$ is calculated considering that an ACK packet
%transmission at the AP corresponds to the generation of $b$ data
%packets in the uplink.

\begin{equation}\label{eq:CTflow_del}
CT_{flow,del} \cong (\frac{n_{up}}{b}+n_{down}) \cdot CT_{AP}
\end{equation}
\begin{equation}\label{eq:w_lim_del}
W_{lim}= \frac{2\cdot
LD}{CT_{flow,del}}+\frac{BS_{AP}}{n_{up}/b+n_{down}}
\end{equation}

\subsection{Calculating $CT_{AP}$}\label{sec:CTAP}

We are interested in the case when there are two active
(saturated) stations (as the AP contends with one station at a
time). The average cycle time in this scenario can easily be
calculated using the model in \cite{Medepalli05}. In our case, the
AP sends the TCP ACK packets of the uplink TCP connections and the
TCP data packets of the downlink TCP connections which contend
with the TCP ACK packets of the downlink TCP connections and the
TCP data packets of the uplink connections that are generated at
the stations. Note that the cycle time varies according to the
packet size of contending stations. Then,
\begin{equation}
\label{eq:CTAP} CT_{AP} = \sum_{p_{1} \in S} {\rm
Pr}(p_{AP}=p_{1}) \sum_{p_{2} \in S} {\rm
Pr}(p_{STA}=p_{2})~CT_{p_{1},p_{2}}
\end{equation}

\noindent where $S=\{ACK, DATA\}$ is the set of different types of
packets, ${\rm Pr}(p_{AP}=p_{1})$ is the probability that the AP
is sending a packet of type $p_{1}$, ${\rm Pr}(p_{STA}=p_{2})$ is
the probability that the non-AP station is sending a packet of
type $p_{2}$, and $CT_{p_{1},p_{2}}$ is the average cycle time
when one station is using a packet of type $p_{1}$ and the other
is using a packet type of $p_{2}$. We differentiate between the
data and the ACK packets because the size of the packets thus the
cycle time duration depends on the packet type.

Using simple probability theory, we can calculate ${\rm
Pr}(p_{AP})$ and ${\rm Pr}(p_{STA})$ as follows
\begin{equation}
\label{eq:pAP} \setlength{\nulldelimiterspace}{0pt} {\rm
Pr}(p_{AP}=p_{1}) =
\left\{ \\
\begin{IEEEeqnarraybox}[\relax][c]{ll} \frac{n_{down}}{n_{up}/b+n_{down}}, & ~ {\rm if} ~ p_{1} = DATA \\
\frac{n_{up}/b}{n_{up}/b+n_{down}}, & ~ {\rm if} ~ p_{1}=ACK,
\end{IEEEeqnarraybox}
\right.
\end{equation}
\begin{equation}
\label{eq:pSTA} \setlength{\nulldelimiterspace}{0pt} {\rm
Pr}(p_{STA}=p_{2}) =
\left\{ \\
\begin{IEEEeqnarraybox}[\relax][c]{ll} \frac{n_{up}}{n_{down}/b+n_{up}}, & ~ {\rm if} ~ p_{2} = DATA \\
\frac{n_{down}/b}{n_{down}/b+n_{up}}, & ~ {\rm if} ~ p_{2}=ACK.
\end{IEEEeqnarraybox}
\right.
\end{equation}

\subsection{Fair Congestion Window Assignment (FCWA)}

%We calculate the maximum flow-specific congestion window limit
%$W_{lim}$ that achieves fair TCP access in a wired/wireless
%scenario where the wireless hop is an 802.11 link.
A control block located at the AP can modify the advertised
receiver window field of the ACK packets that are all relayed
through with the $W_{lim}$ value calculated using the proposed
model. Therefore, we call this procedure Fair Congestion Window
Assignment (FCWA).
%In order to calculate $W_{lim}$,
%the control block at the AP needs
%to estimate $LD$ and $b$.
The analysis requires accurate estimations on $LD$ and $b$. The
control block may distinguish among TCP connections via the IP
addresses and the ports they use. An averaging algorithm can be
used to calculate the average time that passes between sending a
data (ACK) packet into the wired link and receiving the ACK (data)
packet which generated by the reception of the former packet
(which is $2\cdot LD$). The TCP header of consecutive ACK packets
may be parsed to figure out the value of $b$. %\footnotemark.
%\footnotetext{Note that the proposed analysis can easily be
%extended for the case when the value of $b$ differs among TCP
%connections.}
It is also worth to note that although the analytical calculation
uses a simple cycle time method in calculating $CT_{AP}$ and
$CT_{flow}$, the AP may use a measurement-based technique rather
than the model-based technique used in this paper.

\subsection{Buffer Sizing}

The proposed analysis can also directly be used for buffer sizing
purposes. The 802.11 vendors may use the proposed method with
statistics of TCP connections and WLAN traffic to decide on a
\textit{good} size of AP buffer that would provide fair TCP
access.
\begin{equation}\label{eq:BS_AP_del}
BS_{AP}= (W_{lim} - \frac{2\cdot LD}{CT_{flow}}) \cdot
(n_{up}/b+n_{down})
\end{equation}

\subsection{Performance Evaluation}\label{subsec:analysis_evaluation}

We validate the analytical results obtained from the proposed
model via comparing them with the simulation results obtained from
ns-2 \cite{ns2}. We obtained $W_{lim}$ via simulations in such a
way that increasing the TCP congestion window limit of TCP
connections by one results in a packet loss ratio larger than
$1\%$ at the AP buffer.

As previously stated, the network topology is such that each
wireless station initiates a connection with a wired station and
where the traffic is relayed to/from the wired network through the
AP. The TCP traffic uses a File Transfer Protocol (FTP) agent
which models bulk data transfer. TCP NewReno with its default
parameters in ns-2 is used. All the stations have 802.11g PHY
\cite{802.11g} with 54 Mbps and 6 Mbps as the data and the basic
rate respectively. The wired link data rate is 100 Mbps. The
default DCF MAC parameters are used \cite{802.11}. The packet size
is 1500 bytes for all flows. The MAC buffer size at the stations
and the AP is set to 100 packets.

\begin{figure}[t]
\centering \includegraphics[width =
1.0\linewidth]{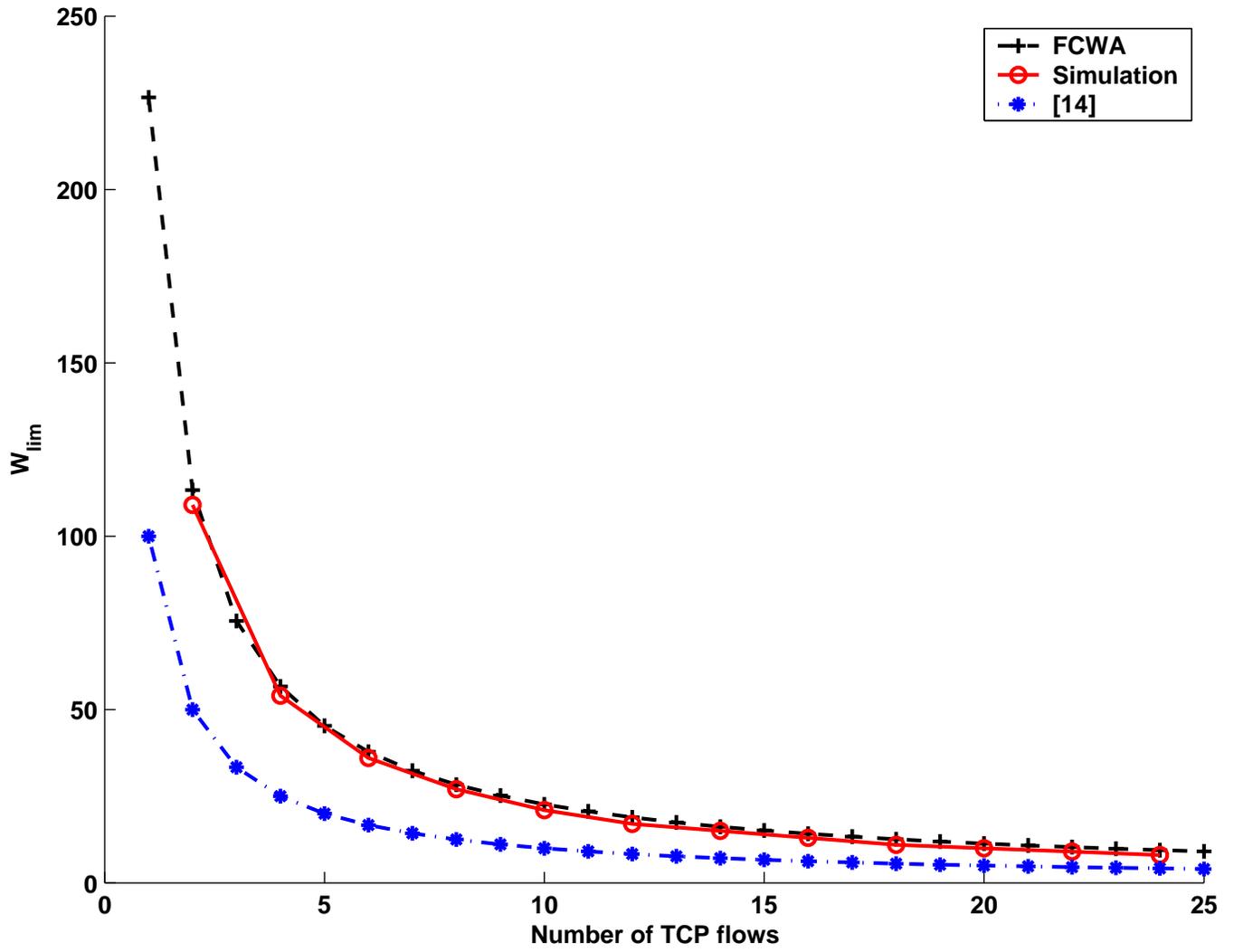} \caption{Congestion window limits
calculated by FCWA, the simulation, and \cite{Pilosof03}.}
\label{fig:cwlim}
\end{figure}

In the first set of experiments, we set the wired link delay of
each connection to 50 ms. Each TCP data packet is acknowledged by
an ACK packet ($b=1$). In Fig.~\ref{fig:cwlim}, we compare the
estimation of (\ref{eq:w_lim}) on the congestion window limit with
the values obtained from the simulation results and the proposed
method of \cite{Pilosof03} for increasing number of TCP
connections. The number of upload flows is equal to the number of
download flows.
%\footnotemark.\footnotetext{Note that for the specific
%scenario, it does not matter whether the TCP connection is in the
%downlink or uplink. Both (\ref{eq:CTflow}) and (\ref{eq:w_lim})
%depend on the total number of flows.}
As Fig.~\ref{fig:cwlim} implies, the analytical results for FCWA
and the simulation results are well in accordance. The analysis in
\cite{Pilosof03} calculates the congestion window limit by
$BS_{AP}/(n_{up}+n_{down})$ and underestimates the actual fair TCP
congestion window limit.
%We observe that in the case of large number of flows, $W_{lim}$
%values that prevent the AP buffer from overflowing are too small
%to be practical. Although small $W_{lim}$ provides fair transport
%layer access, this also results in considerably low per flow
%throughput.

\begin{figure}[t]
\centering \includegraphics[width =
1.0\linewidth]{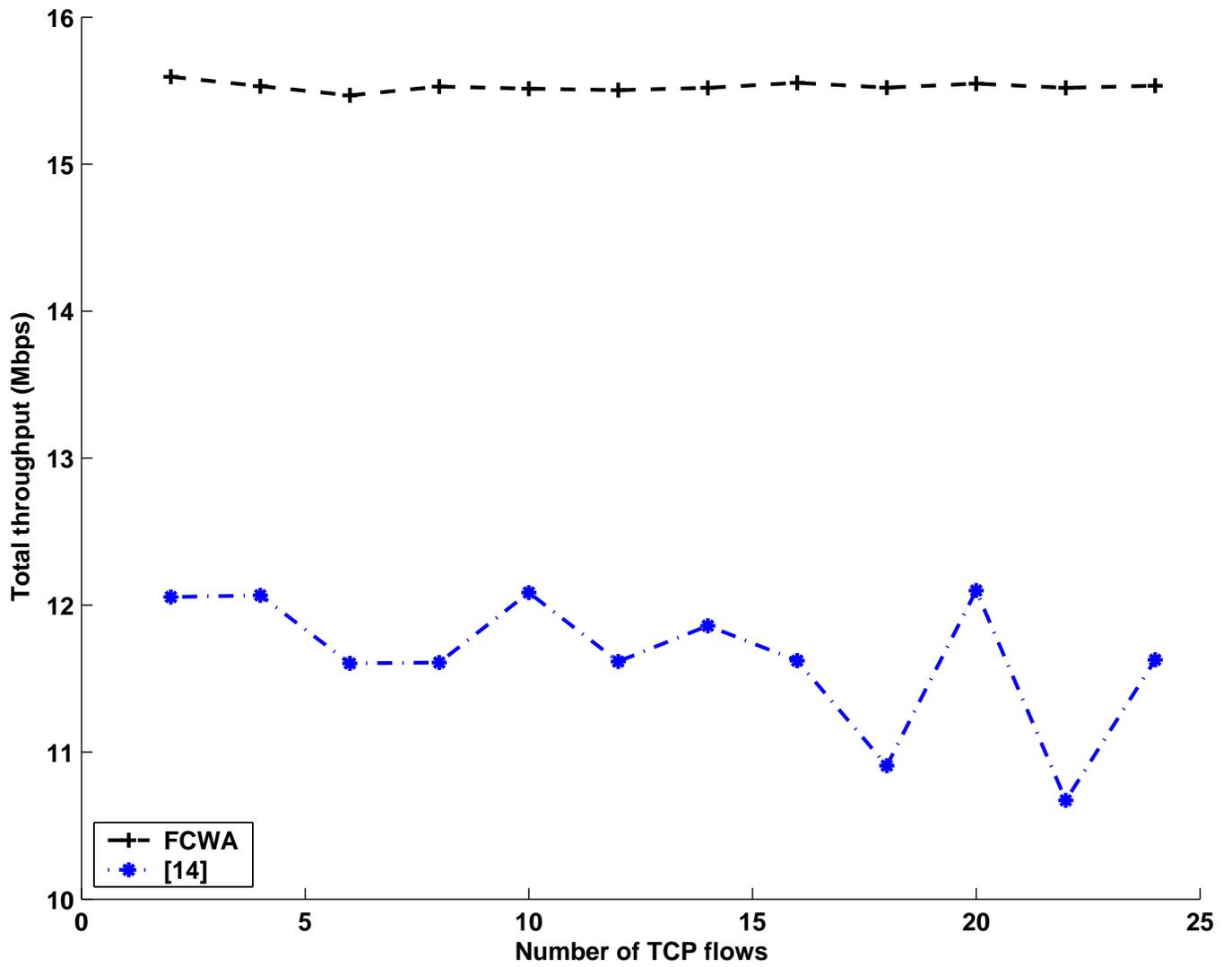} \caption{Total throughput of the
system when the TCP connections employ analytically calculated
congestion window limits by FCWA and \cite{Pilosof03}.}
\label{fig:simplethput}
\end{figure}

The total throughput of the system when the TCP connections employ
analytically calculated congestion window limits in simulation for
increasing number of TCP connections is shown in
Fig.~\ref{fig:simplethput}. As the comparison with
\cite{Pilosof03} reveals, the congestion window limits calculated
via FCWA result in approximately 35\% - 50\% higher channel
utilization for the specific scenario. Although the corresponding
results are not displayed, both methods achieve perfect fairness
in terms of per-connection FTP throughput (Jain's fairness index
\cite{Jain91}, $f> 0.9999$ where 1 shows perfect fairness).

\begin{figure}[t]
\centering \includegraphics[width =
1.0\linewidth]{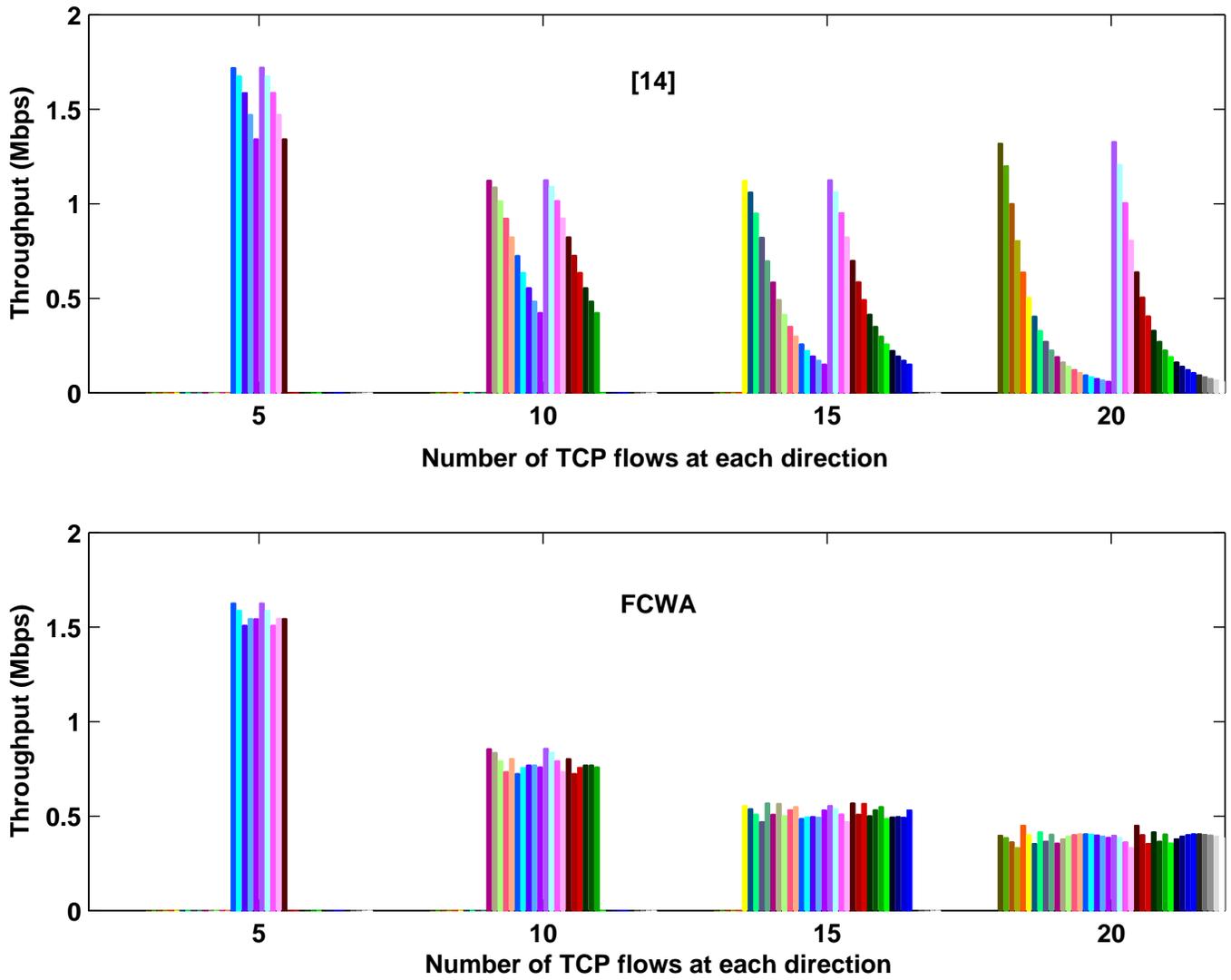} \caption{Individual
throughput of TCP connections when the TCP connections with
different wired link delays employ the analytically calculated
congestion window limits by FCWA and \cite{Pilosof03}.}
\label{fig:diffRTTthput}
\end{figure}

In the second set of experiments, we consider a scenario where
wired link delays ($LD$) among TCP connections differ. First TCP
connection has a 1 ms wired link delay, and $n^{th}$ connection
has $n$ ms larger wired link delay than $(n-1)^{th}$ connection.
%We first calculate the congestion window limits via the proposed
%analysis in this paper and in \cite{Pilosof03} for varying number
%of uplink and downlink connections, and then employ the
%corresponding values in simulation.
Fig. \ref{fig:diffRTTthput} shows the individual throughput for
each TCP connection for FCWA and \cite{Pilosof03} for 4 different
scenarios. In Fig. \ref{fig:diffRTTthput}, for any scenario, the
first half are upload flows and the rest are download flows. As
the results present, the congestion window limits calculated by
the proposed model maintains fair access even in the case of
varying wired link delays. On the other hand, the method proposed
in \cite{Pilosof03} fails to do so.

\begin{figure}[t]
\centering \includegraphics[width =
1.0\linewidth]{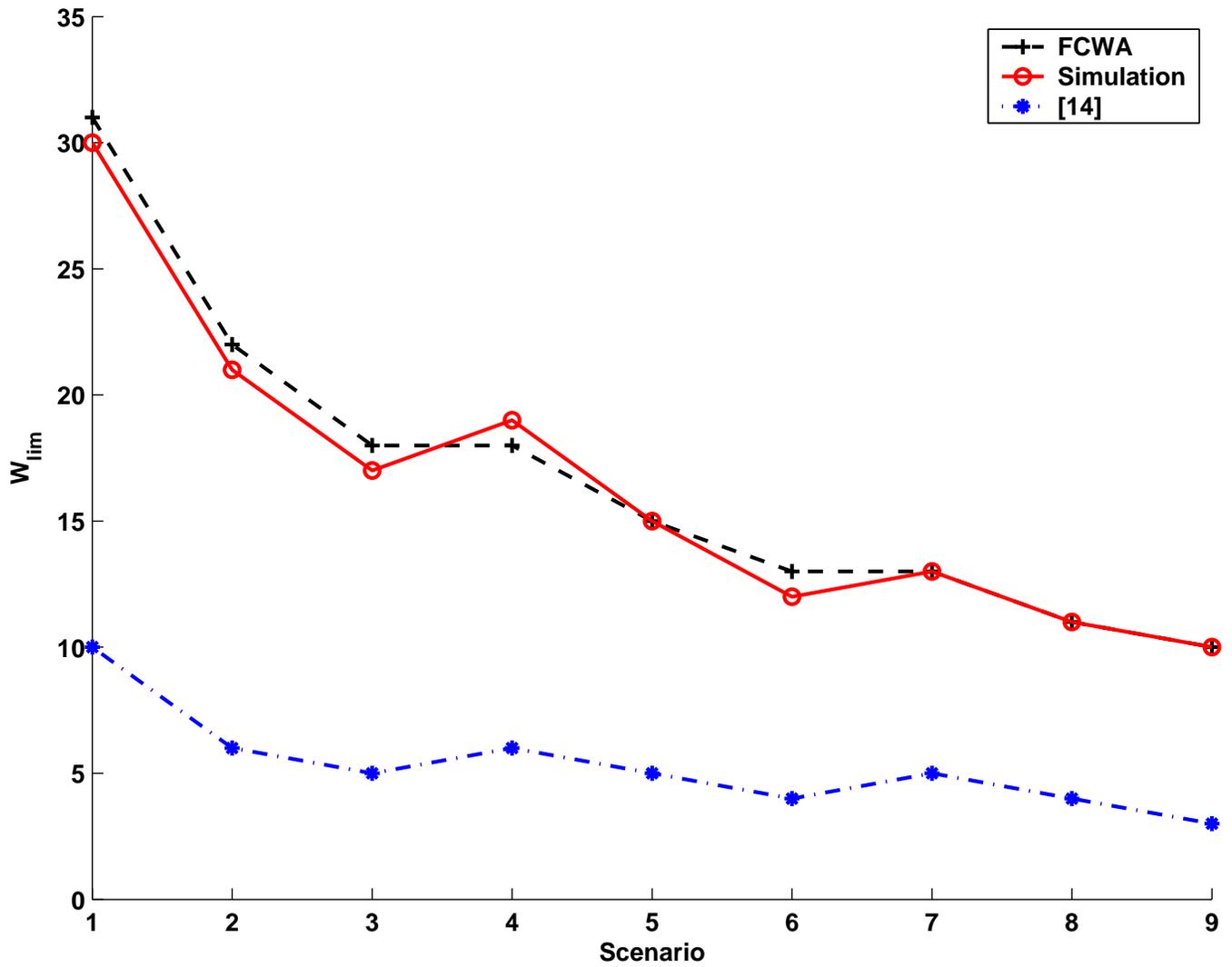} \caption{Congestion window
limits calculated by FCWA, the simulation, and \cite{Pilosof03}
when TCP connections employ delayed ACK mechanism with $b=2$.}
\label{fig:cwlim_del}
\end{figure}

\begin{figure}[t]
\centering \includegraphics[width =
1.0\linewidth]{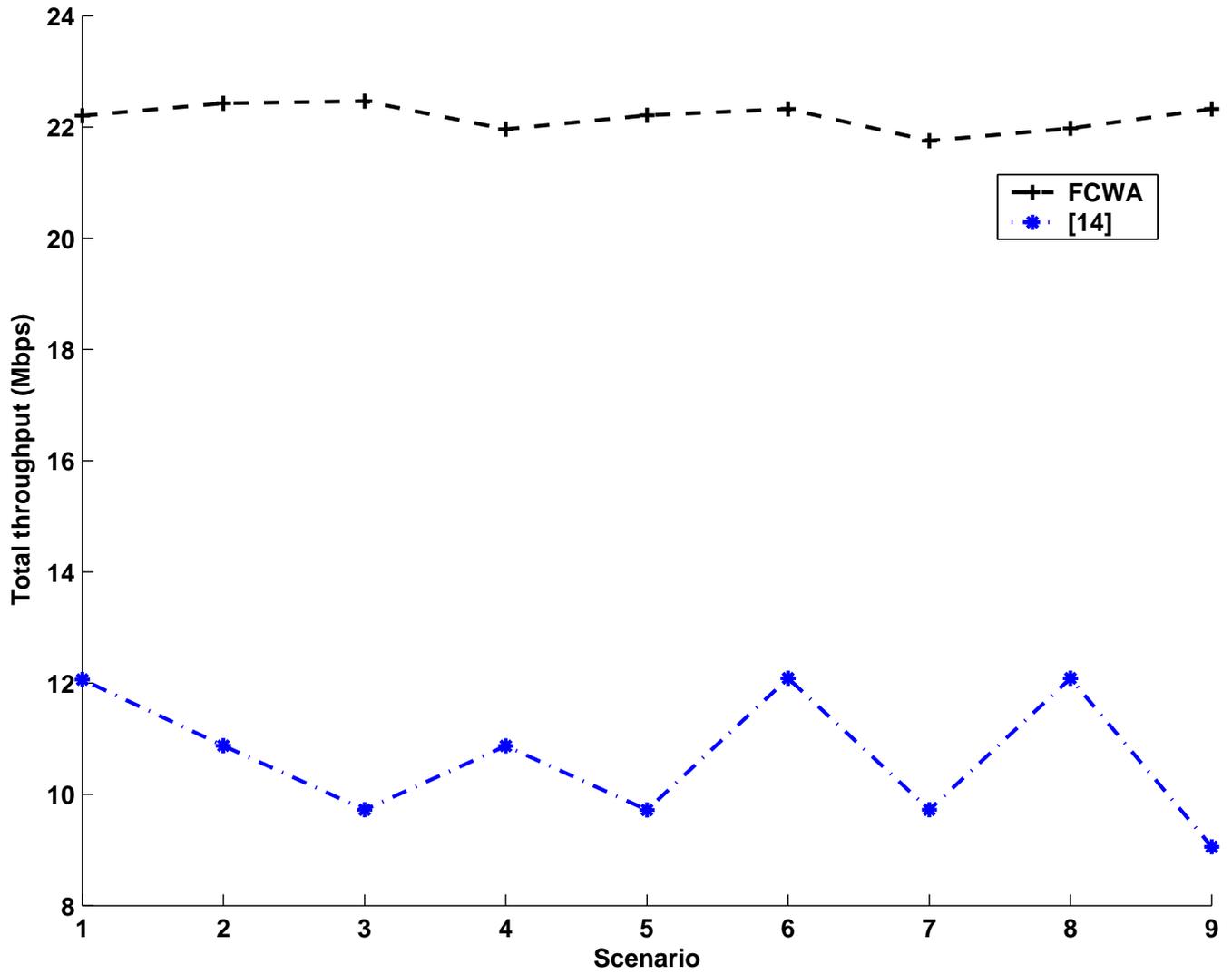} \caption{Total throughput of
the system when the TCP connections employ analytically calculated
congestion window limits by FCWA and \cite{Pilosof03} in the case
delayed ACK mechanism is used ($b=2$).} \label{fig:delayedthput}
\end{figure}

In the third set of experiments, we consider a scenario where the
TCP connections use the delayed ACK mechanism with $b=2$. We
consider 9 different scenarios where in each scenario the number
of uplink and downlink TCP connections varies. In the first three
scenarios, the number of downlink flows is set to 5 and the number
of uplink flows is varied among 5, 10, and 15, respectively.
Varying the number of uplink flows in the same range, the next
three scenarios use 10, and the following three scenarios use 15
downlink flows. In Fig.~\ref{fig:cwlim_del}, we compare the
estimation of (\ref{eq:w_lim_del}) on the congestion window limit
with the values obtained from the simulation results and the
proposed method of \cite{Pilosof03}. The analytical results for
the proposed model and the simulation results are well in
accordance.The total throughput of the system when the TCP
connections employ the analytically calculated congestion window
limits is shown in Fig.~\ref{fig:delayedthput}. As the comparison
with \cite{Pilosof03} reveals, the congestion window limits
calculated via our method result in approximately 90\% - 105\%
higher channel utilization. Although the corresponding results are
not presented, the congestion window limits calculated by both the
proposed method and the method of \cite{Pilosof03} achieve
perfectly fair resource allocation in terms of throughput (Jain's
fairness index, $f > 0.998$). On the other hand, the proposed FCWA
method results in a significantly higher channel utilization.

%As Fig.~\ref{fig:fairness_vs_cw4_dif_flow_numbers} and
%Fig.~\ref{fig:thput_vs_cw4_dif_flow_numbers} show the maximum
%congestion window sizes larger than $W_{lim}$ may also provide
%fair access and higher channel utilization (on average) up to a
%degree. The main characteristic of the total throughput curves is
%that the throughput increases with increasing maximum congestion
%window size for comparably small congestion windows. Then, it
%stays constant for a while. After the maximum congestion window
%size value gets larger than $W_{lim}$, it starts to increase
%again. This is due to cumulative nature of TCP ACKs. Although some
%of the TCP ACKs are lost at the AP buffer, the system maintain
%fairness and higher channel utilization until the ACK loss becomes
%severe. Observing that some ACK losses do not affect fairness and
%even increase the total throughput, we propose filtering and
%scheduling the ACKs in an intelligent way.

\section{Link Layer Access Control Block}\label{sec:proposed}

As illustrated in Section \ref{subsec:problem}, unfair access
problem originates from the uplink/downlink bandwidth asymmetry in
the 802.11 BSS. As our analysis in Section \ref{sec:analysis}
shows, fair access can be achieved if the congestion window limits
of the downlink and the uplink TCP sources are set regarding the
network bandwidth so that no packet drops occur at the AP buffer.
Actually, for the default DCF scenario when only downlink flows
are present, the data packet drops at the AP buffer implicitly
throttles the downlink TCP sources effectively (i.e., TCP access
is fair among downlink flows in this case as also we present in
\cite{Keceli08_fairTCP_trep} via simulations). Conversely, the
coexistence of uplink flows shuts down the downlink as some uplink
flows are fortunate enough to reach a high congestion window by
making use of the cumulative property of TCP ACKs. The TCP ACKs of
uplink flows occupy most of the AP buffer which results in data
packet drops for downlink flows.

%As illustrated in Section \ref{subsec:problem}, downlink TCP flows
%may starve in terms of throughput if a few uplink TCP flows
%coexist in an 802.11 infrastructure BSS. This is a direct result of the effect of a TCP ACK packet loss
%being different from the effect of a TCP data packet loss. Using
%the cumulative nature of TCP ACK packets, most uplink flows can
%sustain high throughput. On the other hand, downlink flows
%experience frequent timeouts and are shut out.

These observations motivate the approach of our novel idea:
Prioritize TCP data packets of downlink flows over TCP ACK packets
of uplink flows at the AP MAC buffer. We design a novel link layer
access control block which employs an ACK Congestion Control and
Filtering (ACCF) scheme.
%The appropriate prioritization
%between the downlink data and the uplink ACKs which leads to fair
%access provisioning is achieved by adapting the parameters of the
%proposed ACK filtering and scheduling scheme regarding the
%measurements on downlink traffic.
The proposed ACCF scheme delays the TCP ACK packets of uplink
flows (using a separate control block buffer) regarding the
measured average packet interarrival time of the downlink TCP data
packets. In other words, the downlink data to uplink ACK
prioritization ratio is quantified by means of estimating what the
uplink ACK transmission rate should be for the given average
downlink TCP data transmission rate. The rationale behind the
proposed method is sending the TCP ACKs of uplink connections only
as often as the TCP data of downlink connections are sent.

%The proposed link layer access control block lies above the MAC
%layer at the AP. The link layer access control block adaptively
%decides the time an ACK packet to be sent down to the AP MAC
%buffer. Any packet that enters the AP MAC queue uses DCF rules to
%access the channel. Therefore, the proposed solution does not
%require any changes in the 802.11 standard, nor any enhancement at
%the stations.

The proposed ACCF algorithm uses the cumulative property of TCP
ACKs by employing ACK filtering. If another ACK packet of flow $i$
is received while there is an ACK packet of flow $i$ in the
control block buffer, the previous ACK in the buffer is replaced
with the new one. Our rationale behind introducing ACK filtering
is to reduce the number of ACK packets transmitted by the AP. This
creates more room in the AP buffer for TCP data packets of
downlink flows (which in turn decreases TCP data packet loss
ratio). Moreover, filtering ACK packets also slows the growth rate
of TCP congestion windows of uplink flows (since the TCP senders
receive less frequent ACK packets) which further limits the share
of the uplink bandwidth.

%Note that this functionality results in downlink TCP data packets
%enjoying higher priority when compared to uplink TCP ACK packets.

%The proposed link layer access control block runs a simple TCP ACK
%congestion control and filtering algorithm. The proposed algorithm
%adaptively determines the uplink TCP ACK to be enqueued to the IFQ
%according to the estimates of the average interarrival time of
%downlink TCP data packets.  Enjoying the higher priority, the TCP
%data packets of downlink flows are not delayed and filtered by the
%proposed access control block and directly enqueued to the IFQ.

We define the following notation for the description of the
algorithm provided in the sequel. Let $num_{cum,i}$ be the current
number of accumulated ACKs for flow $i$ in the access control
block buffer.
%Let $AvgDataInt$ be the average data packet
%interarrival time measured at the AP for all downlink TCP flows.
Let $t_{buf,i}$ denote the total time that has passed since the
last TCP ACK for flow $i$ has been sent to the MAC queue. Let
$\beta$ be a constant weighing factor and $\gamma$ be a variable
weighing factor which is a function of $num_{cum,i}$. Let
$AvgInt_{i}$ be the measured average packet interarrival time for
flow $i$\footnotemark\footnotetext{$AvgInt_{i}$ denotes the
average TCP data packet interarrival time if flow $i$ is a
downlink TCP flow and the average TCP ACK packet interarrival time
if flow $i$ is an uplink TCP flow. $AvgInt_{i}$ can be calculated
by employing simple averaging methods (such as exponential moving
averaging that we have employed for uplink measurements in
\cite{Keceli07_ICC}) on periodic measurements results.}. Let
$AvgDataInt$ be the average downlink data interarrival duration
which we use on deciding how frequent the ACKs of uplink flows
should be sent down to the MAC queue for transmission. We
calculate $AvgDataInt$ by taking the mean of $AvgInt$ of the
downlink flows with $AvgInt<\alpha\cdot\min({AvgInt_{j}:\forall
j~in~downlink})$ where $1<\alpha$ is a constant. Note that this
averaging calculation excludes the TCP sources with packet
interarrivals higher than a threshold (as quantified by $\alpha$)
in order to prevent slow downlink flows limiting the frequency of
uplink ACKs, therefore the uplink bandwidth unnecessarily.

According to the proposed ACCF algorithm, the TCP ACKs are
scheduled for transmission (sent down to the MAC queue) such that
the average per-flow ACK rate does not exceed the average per-flow
TCP downlink packet rate. Using this idea, we quantify the control
queue buffering time for each ACK packet of uplink flow $i$ as
$D_{i}=\gamma \cdot num_{cum,i} \cdot AvgDataInt - t_{buf,i}$. The
rationale behind this equation is as follows.
\begin{itemize}
\item We consider the cumulative number of ACK packets that the
currently buffered ACK packet represents. The transmission of an
accumulated ACK packet is expected to trigger the generation of
$num_{cum,i}$ data packets in the uplink. Therefore, any
accumulated TCP ACK packet is delayed until that many TCP downlink
data transmissions can be made on average ($num_{cum,i} \cdot
AvgDataInt$).
\item If a
few consecutive timeouts are experienced when the TCP congestion
window is small, the uplink TCP flow may hardly recover, and
consequently may suffer from low throughput (as we also observed
via simulations). Therefore, we introduce an adaptive weighing
factor $\gamma_{min}\leq \gamma \leq 1$ in the minimum buffering
duration. We use the value of $num_{cum,i}$ as an indication of
the current size of the TCP congestion window of the corresponding
flow. %\footnotemark.
%Our reasoning behind this will be described later in this
%section.
The value of $\gamma$ is set smaller than 1 when $num_{cum,i}$ is
smaller than a threshold, $num_{thresh}$. The idea is to prevent
longer delays at the control block buffer thus possible timeouts
at the TCP agent at the station if the uplink TCP connection is
expected to have a small instantaneous congestion window (e.g., a
recently initiated TCP connection).
\item
%The delay to be introduced to any accumulated ACK is
%normalized by the total time that has passed since the last TCP
%ACK for flow $i$ has been sent to the MAC queue ($t_{buf,i}$).
%The proposed delay limit results in approximately scheduling the
%transmission of a TCP ACK packet acknowledging $num_{cum,i}$ data
%packets every interval with duration equal to $\gamma\cdot
%num_{cum,i} \cdot AvgDataInt$.
We substract $t_{buf,i}$ from $\gamma \cdot num_{cum,i} \cdot
AvgDataInt$ in order to make
%When TCP congestion
%window size is sufficiently large, i.e.,
%$num_{cum,i}>num_{thresh}$, $\gamma$ is 1. Therefore, according to
%the proposed limit,
the duration of the interval between two consecutive ACKs sent
down to the MAC buffer approximately equal to $num_{cum,i} \cdot
AvgDataInt$ (in the case $num_{cum,i}>num_{thresh}$).
%Each ACK
%cumulatively acknowledges $num_{cum,i}$ data packets. This
%approximately results in an average uplink TCP data transmission
%interval equal to an average downlink data transmission interval
%$AvgDataInt$ (which is required for achieving uplink/downlink fair
%TCP access).
\end{itemize}

%\footnotetext{Due to the previously mentioned minimum ACK delay
%based on average interarrival time ($\beta \cdot AvgInt_{i}$), the
%control block sends one ACK for flow $i$ per congestion window on
%the average. Therefore, the cumulated number of ACKs at the AP is
%usually a good indication on the actual window size.}

%We consider the total time passed since the arrival of the first
%ACK packet (which is not an accumulated ACK packet) in the control
%queue.

As we have also observed via simulations, the ACK filtering scheme
makes the ACK arrivals to the AP queue bursty
\cite{Balakrishnan99}. For an arbitrary uplink flow, this behavior
corresponds to alternating idle times with no packet arrivals and
active times consisting of a bunch of highly frequent ACK arrivals
to the AP queue. This bursty behavior may result in
$t_{buf,i}>\gamma \cdot num_{cum,i} \cdot AvgDataInt$ (probably
when the corresponding idle duration is long), therefore
$D_{i}<0$, especially for the first few ACK arrivals at the AP
queue following an idle time for the corresponding flow. Note that
the case of $D_{i}<0$ actually translates into the case of the ACK
being already due for transmission. In this case, our design takes
one of the two alternative actions regarding the value of $D_{i}$
as follows.
\begin{itemize}
\item $D_{i}+\beta \cdot AvgInt_{i}<0$: This serves as an
indication of the last ACK pass to the MAC queue having been done
probably within the previous burst. Although the ACK transmission
is due ($D_{i}<0$), an immediate pass to the MAC queue punishes
uplink throughput unnecessarily as $t_{buf,i}/num_{cum,i}$ (which
is an indication of average data transmission interval for uplink
flow $i$) is much larger than $AvgDataInt$. In this case, the ACK
packet of flow $i$ is delayed for the duration equal to
$D'_{i}=\beta \cdot AvgInt_{i}$ in the control block queue
(counting on the high probability of further ACK arrivals in the
current burst).
%Introducing such a delay paves the way for the ACK filtering
%algorithm.
Our intuition behind the calculation of $D'_{i}$ is the
possibility of the next ACK of the same flow arriving possibly in
an average interarrival time $AvgInt_{i}$. We also introduce the
constant weighing factor $\beta>1$ in order to compensate for the
potential variance of the \textit{instantaneous} ACK interarrival
time. A new ACK arrival will probably decrease
$t_{buf,i}/num_{cum,i}$ taking it closer to $AvgDataInt$.
\item If $D_{i}+\beta \cdot AvgInt_{i}>0$, the relaying ACK packet is sent
down to the MAC queue as the ACK is already due for transmission
and $t_{buf,i}/num_{cum,i}$ is close to $AvgDataInt$.
\end{itemize}

%Combining both limits we proposed on the duration of TCP ACK
%delay, we can define the simple rule on how long the relaying ACK
%packet waits in the control buffer before it is enqueued to the
%MAC queue for transmission
%\begin{equation}\label{eq:Di}
%D_{i} = {\rm max}(\beta \cdot AvgInt_{i}, \gamma \cdot num_{cum,i}
%\cdot AvgDataInt - t_{buf,i}).
%\end{equation}

As previously stated, if a new TCP ACK packet arrives before the
timer that is initially set to $D_{i}$ (or $D'_{i}$) at the
arrival of previous ACK expires, the new ACK replaces the
previously buffered ACK. The link layer access control block
parses the TCP header to calculate $num_{cum,i}$ and restarts the
timer with the new $D_{i}$ (or $D'_{i}$) for the accumulated ACK.
When the timer expires, the TCP ACK is sent down to the MAC queue
and both $t_{buf,i}$ and $num_{cum,i}$ are reset to 0.

%To sum up, the ideas behind (\ref{eq:Di}) are very simple:
%\textit{i)} delay the ACK for an average interarrival time to find
%out if another ACK is coming and \textit{ii)} do not send the ACK
%down to the MAC buffer for transmission if this can create an
%uplink data transmission rate larger than measured data downlink
%transmission rate.

%\footnotetext{Any ACK packet is delayed at least the average ACK
%interarrival time. Also, a weighing factor $\beta>1$ is
%introduced. $\beta \cdot AvgInt_{i}$. Moreover, the second term in
%the comparison is usually larger than the first. As a result of
%all, it is highly probable that any TCP ACK in the AP is delayed
%long enough so that all uplink TCP data of the current congestion
%window is on flight.}

%A number of issues are known about ACK filtering.
ACK filtering
may slow down the congestion window growth rate, negatively impact
the performance during loss recovery and slow start, and increase
the round trip time \cite{Balakrishnan97}. On the other hand,
since our idea is trying to slow down uplink TCP flows in order to
prioritize downlink TCP flows, most of these issues do not
negatively affect fairness and overall channel utilization. Still,
the proposed algorithm does not filter the TCP ACKs with flags set
such as duplicate ACKs which are directly enqueued to the MAC
queue.

%Moreover, if a flow is detected to be newly started, the first few
%packets are not filtered as well. This approach partially combats
%the slow congestion window growth problem during the slow start
%phase.

The proposed ACCF algorithm introduces a number of configurable
variables. As pointed out in Section \ref{subsec:simulations}, we
decided the values for these variables through extensive
simulations.

\subsection{Fairness Measure} \label{sec:fairness_measure}

Most of the studies in the literature quantify the fairness by
employing Jain's fairness index \cite{Jain91} or providing the
ratio of the throughput achieved by individual or all flows in the
specific directions. On the other hand, such measures have the
implicit assumption of each flow or station demanding
asymptotically high bandwidth (i.e., in saturation and having
always a packet ready for transmission). As these measures
quantify, a perfectly fair access translates into each flow or
station receiving an equal bandwidth. On the other hand, in a
practical scenario of flows with finite and different bandwidth
requirements (i.e., some stations in nonsaturation and
experiencing frequent idle times with no packets to transmit),
these measures cannot directly be used to quantify the fairness of
the system.

We define the fair access in a scenario where flows with different
bandwidth requirements coexist as follows.
\begin{itemize}
\item The flows with total bandwidth requirement lower than the fair per-flow channel
capacity in the specific direction receive the necessary
bandwidth.
%so that no packet drops occur at the MAC queue of the
%station (for uplink) or the AP (for downlink).
\item The flows with total bandwidth requirement higher than the fair per-flow
channel capacity receive an equal bandwidth.
% so
%that an equal number of transmit opportunities are provided to the
%stations (for uplink) or the AP (for downlink to serve a specific
%station).
\end{itemize}

In order to quantify fair access, we propose to use the MAC queue
packet loss rate (a packet loss rate of 0 for all flows
corresponds to fair access) for the latter together with the
comparison on channel access rate (equal channel access rate
corresponds to fair access) for the former. Note that the latter
can employ Jain's fairness index, $f$, which is defined in
\cite{Jain91} as follows: if there are $n$ concurrent connections
in the network and the throughput achieved by connection $i$ is
equal to $x_{i}$, $1 \leq i \leq n$, then
\begin{equation}
f =
\frac{\left(\sum_{i=1}^{n}x_{i}\right)^{2}}{n\sum_{i=1}^{n}x_{i}^{2}}.
\end{equation}

%The fairness index lies between 0 and 1. In a fair scenario, every
%saturated station gets an equal throughput ($f=1$) and every
%nonsaturated station achieves a packet loss rate of 0 at the MAC
%queue.

\subsection{Performance Evaluation}\label{subsec:simulations}

We implemented the proposed link layer control access block
employing ACCF in ns-2 \cite{ns2}.
%In this section, we present the simulation results on the
%performance of the network in terms of fairness and throughput.
%Our main performance metric is the widely used fairness index
%\cite{Jain91}. The fairness index, $f$, is defined as follows: if
%there are $K$ concurrent connections in the network and the
%throughput achieved by connection $i$ is equal to $x_{i}$, $1 \leq
%i \leq K$, then
%\begin{equation}
%f =
%\frac{\left(\sum_{i=1}^{K}x_{i}\right)^{2}}{K\sum_{i=1}^{K}x_{i}^{2}}.
%\end{equation}
%
%\noindent The fairness index lies between 0 and 1, 1 being the
%most fair situation where every flow gets equal throughput.
The network topology and the stated parameters in
\ref{subsec:analysis_evaluation} are used. The TCP traffic uses
either a File Transfer Protocol (FTP) agent, which models bulk
data transfer or a Telnel agent, which simulates the behavior of a
user with a terminal emulator or web browser. Unless otherwise
stated, flows are considered to be lasting through the simulation
duration and called long-lived in the sequel. On the other hand,
in some experiments, we also use short-lived TCP flows which
consist of 31 packets and leave the system after all the data is
transferred. The receiver advertised congestion window limits are
set to 42 packets for each flow. Note that the scale on the buffer
size and TCP congestion window limit is inherited from
\cite{Pilosof03}. Although the practical limits may be larger, the
unfairness problem exists as long as the ratio of the buffer size
to the congestion window limit is not arbitrarily large (which is
not the case in practice). We found $\alpha=1.5$, $\beta=2$,
$\gamma_{min}=0.5$, and $num_{thresh}=10$ to be appropriate
through extensive simulations. The simulation duration is 350 ms.

We investigate the system performance when wired link delays
($LD$) differ among TCP connections. The wired link delay of the
first upload or download TCP connection is always set to 10 ms.
Then, any newly generated upload or download TCP connection has a
wired link delay of 2 ms larger than the previous one in the same
direction.

\paragraph{The Basic Scenario} In the first set of experiments,
we generate 3, 5, or 10 upload FTP connections and vary the number
of download FTP connections from 5 to 30. The wireless channel is
assumed to be errorless.

\begin{figure}[t]
\centering \includegraphics[width =
1.0\linewidth]{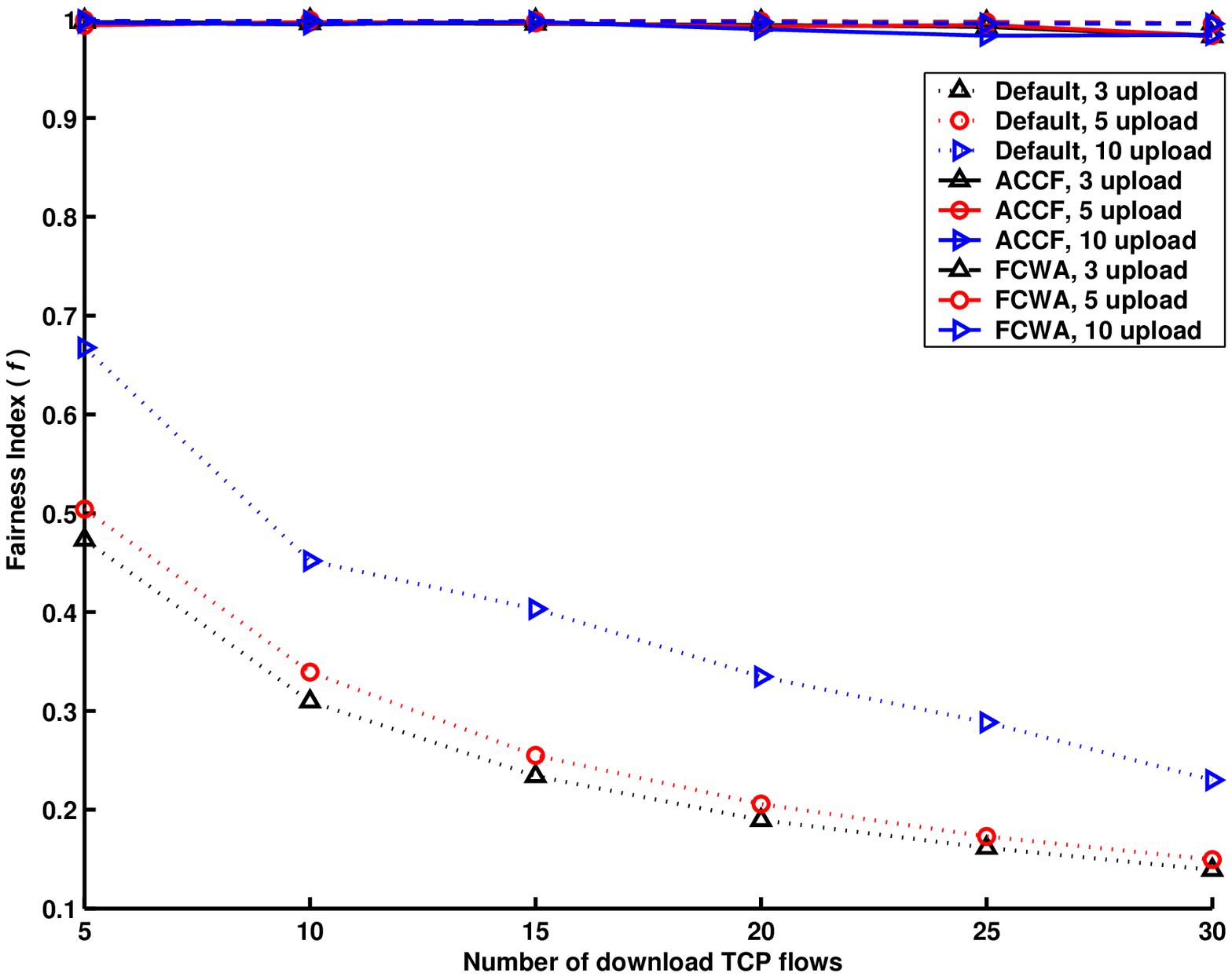} \caption{Fairness index
among all TCP flows when 3, 5, or 10 upload TCP connections are
generated and the number of download TCP connections are varied
from 5 to 30.} \label{fig:simplefairness}
\end{figure}

Fig. \ref{fig:simplefairness} shows the fairness index among all
connections. We compare the default DCF results with the results
obtained when the AP employs the proposed \textit{i)} FCWA or
\textit{ii)} ACCF. As the results imply, with the introduction of
any of the proposed control blocks at the AP, an almost perfect
fair resource allocation can be achieved in both cases.

\begin{figure}[t]
\centering \includegraphics[width =
1.0\linewidth]{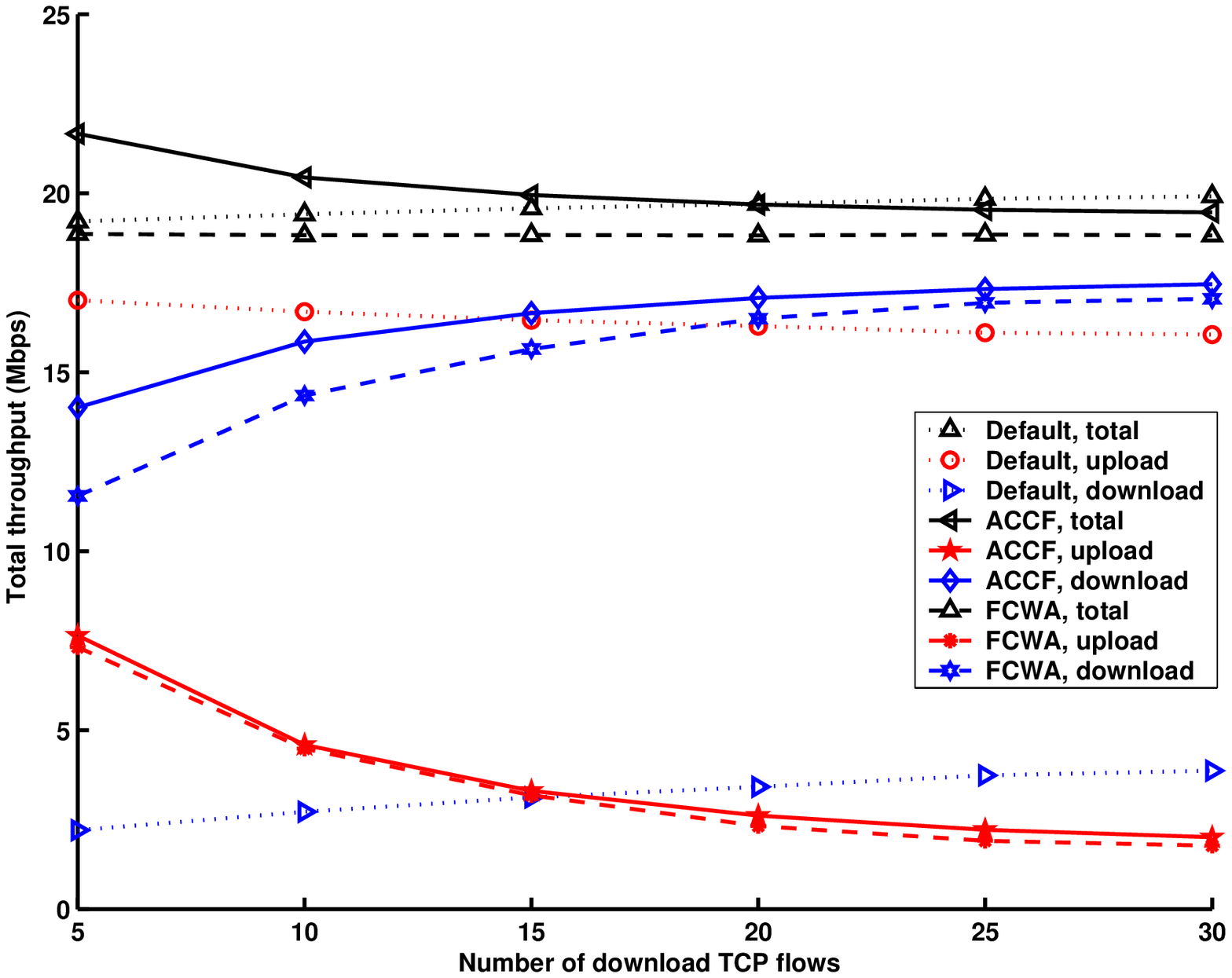} \caption{Throughput of
upload and download TCP connections when 3 upload TCP connections
are generated and the number of download TCP connections are
varied from 5 to 30.} \label{fig:simplethroughput}
\end{figure}

\begin{figure}[t]
\centering \includegraphics[width =
1.0\linewidth]{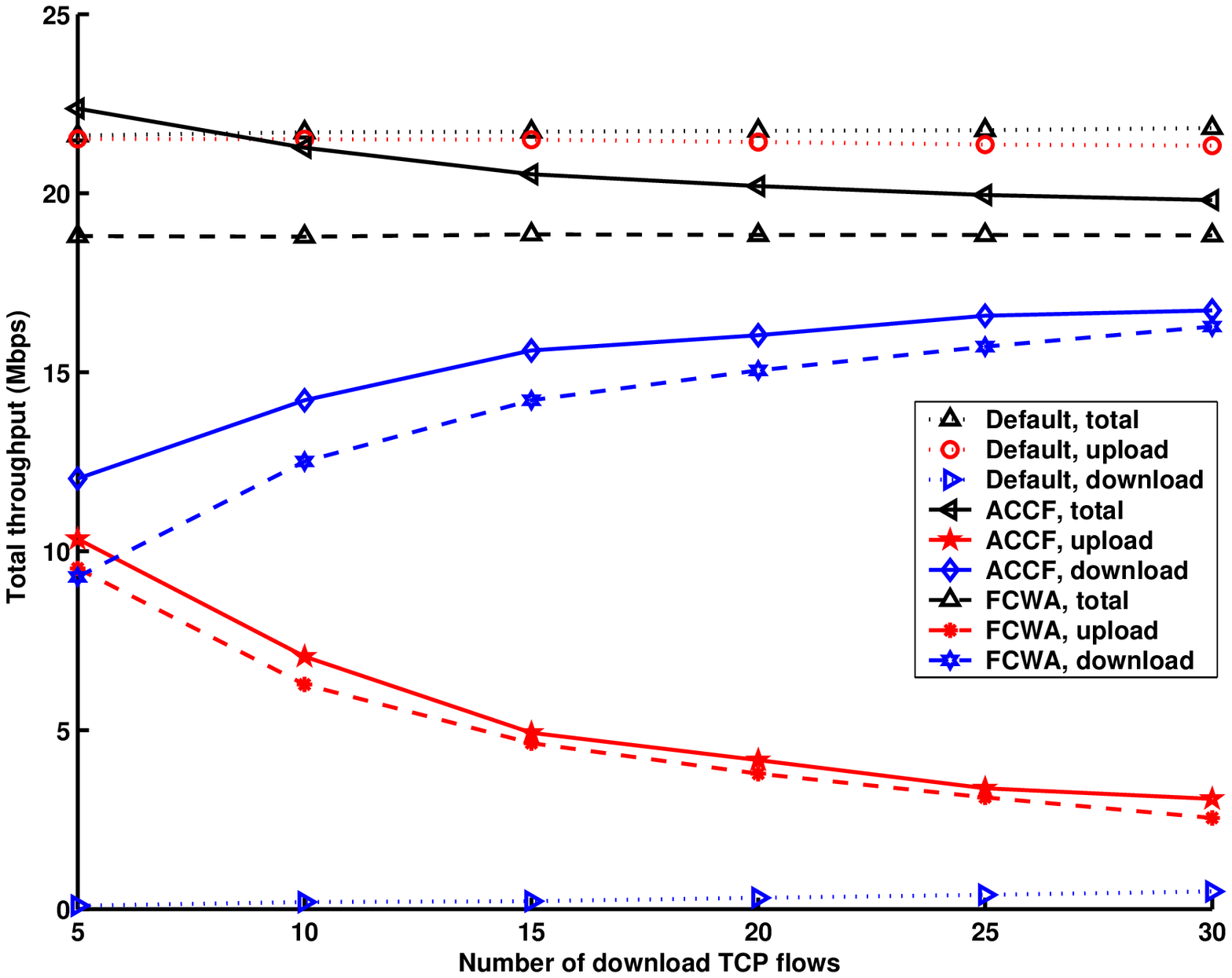} \caption{Throughput of
upload and download TCP connections when 5 upload TCP connections
are generated and the number of download TCP connections are
varied from 5 to 30.} \label{fig:simplethroughput_5}
\end{figure}

\begin{figure}[t]
\centering \includegraphics[width =
1.0\linewidth]{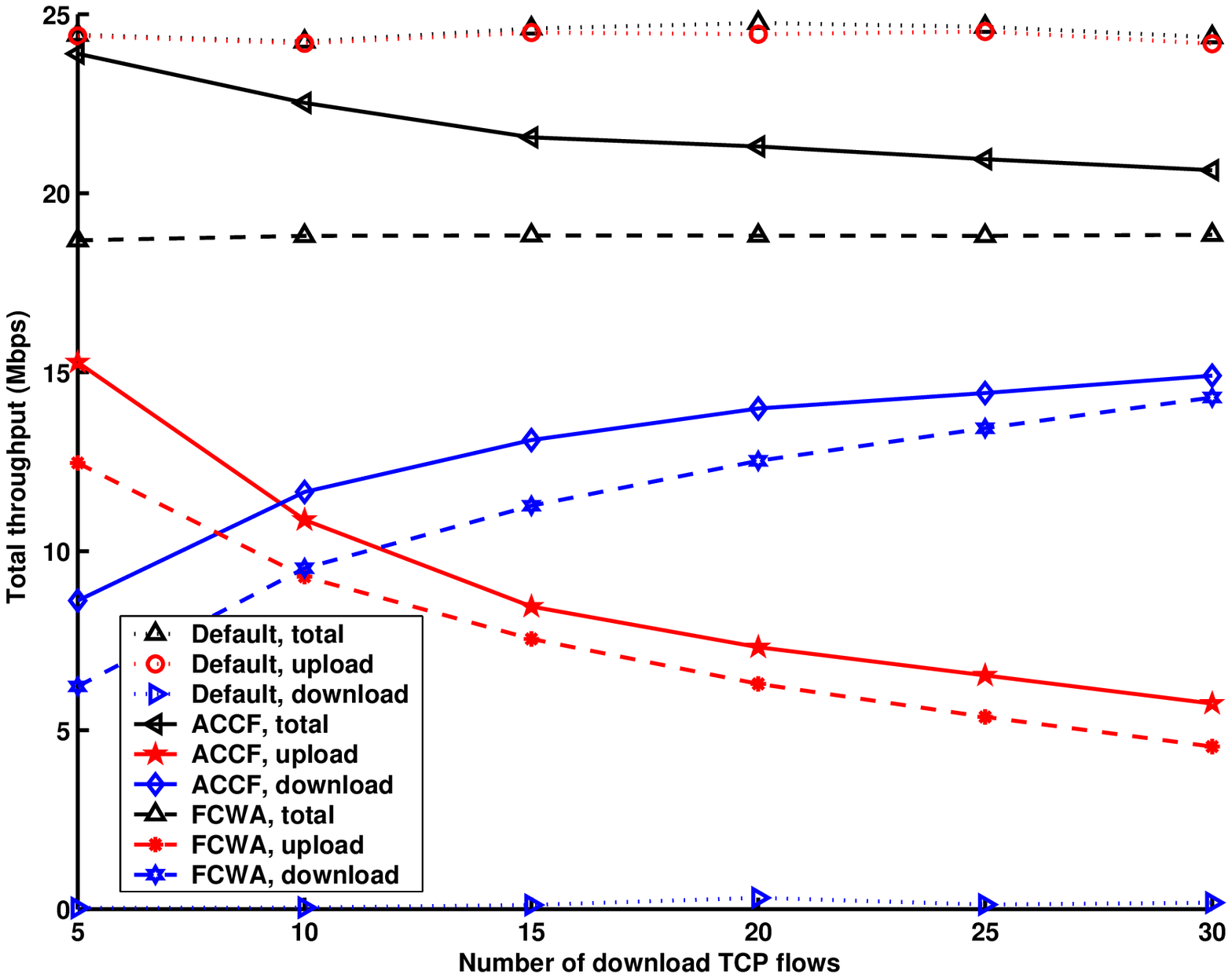} \caption{Throughput
of upload and download TCP connections when 10 upload TCP
connections are generated and the number of download TCP
connections are varied from 5 to 30.}
\label{fig:simplethroughput_10}
\end{figure}

In Fig.~\ref{fig:simplethroughput},
Fig.~\ref{fig:simplethroughput_5}, and
Fig.~\ref{fig:simplethroughput_10}, we plot the uplink, the
downlink, and the total TCP throughput in the infrastructure BSS,
when there are 3, 5, and 10 upload TCP connections, respectively.
As the results show, using the proposed ACCF scheme, the downlink
flows (which starve in the default DCF case) can achieve
reasonable throughput. If we employ FCWA instead, the total
throughput observed is slightly lower. In this case, the proposed
ACCF scheme makes use of the ACK filtering scheme to achieve a
higher channel utilization. The comparison with the performance of
the default DCF algorithm implies that the proposed methods do not
sacrifice channel utilization while providing fair access.

%Note that the total uplink throughput is less than the total
%downlink throughput in Fig. \ref{fig:simplethroughput} for the
%proposed method, since the number of TCP connections in the uplink
%is fewer than the number of TCP connections in the downlink. As
%Fig. \ref{fig:simplefairness} shows, per-flow throughput is fair.

%Fig. \ref{} shows the evolution of TCP congestion window for an
%arbitrarily selected upload and download TCP connection. Although
%not presented here, the other flows present similar behavior. In
%the default case, the TCP congestion window size reaches maximum
%limit easily in the uplink, while the TCP congestion window size
%in the downlink can hardly grow (the download TCP connection
%experiences frequent timeouts). In contrast, the introduction of
%the proposed link layer access control block at the AP constrains
%the window size growth of uplink flows. This allows downlink TCP
%connections to grow reasonably.

\paragraph{Delayed TCP ACKs}

In the second set of experiments, we use a scenario when TCP
connections use the delayed TCP ACK mechanism ($b=2$).
%In the
%delayed TCP ACK mechanism, the TCP receiver acknowledges every $b$
%TCP data packets ($b>1$). We use a typical value (widely used in
%practice), $b=2$, for the specific experiment.
%where every 2 TCP data packets get
%one ACK packet. Legacy ns-2 PHY implements an energy-based PHY
%model \footnotemark{}.

\begin{figure}[t]
\centering \includegraphics[width = 1.0\linewidth]{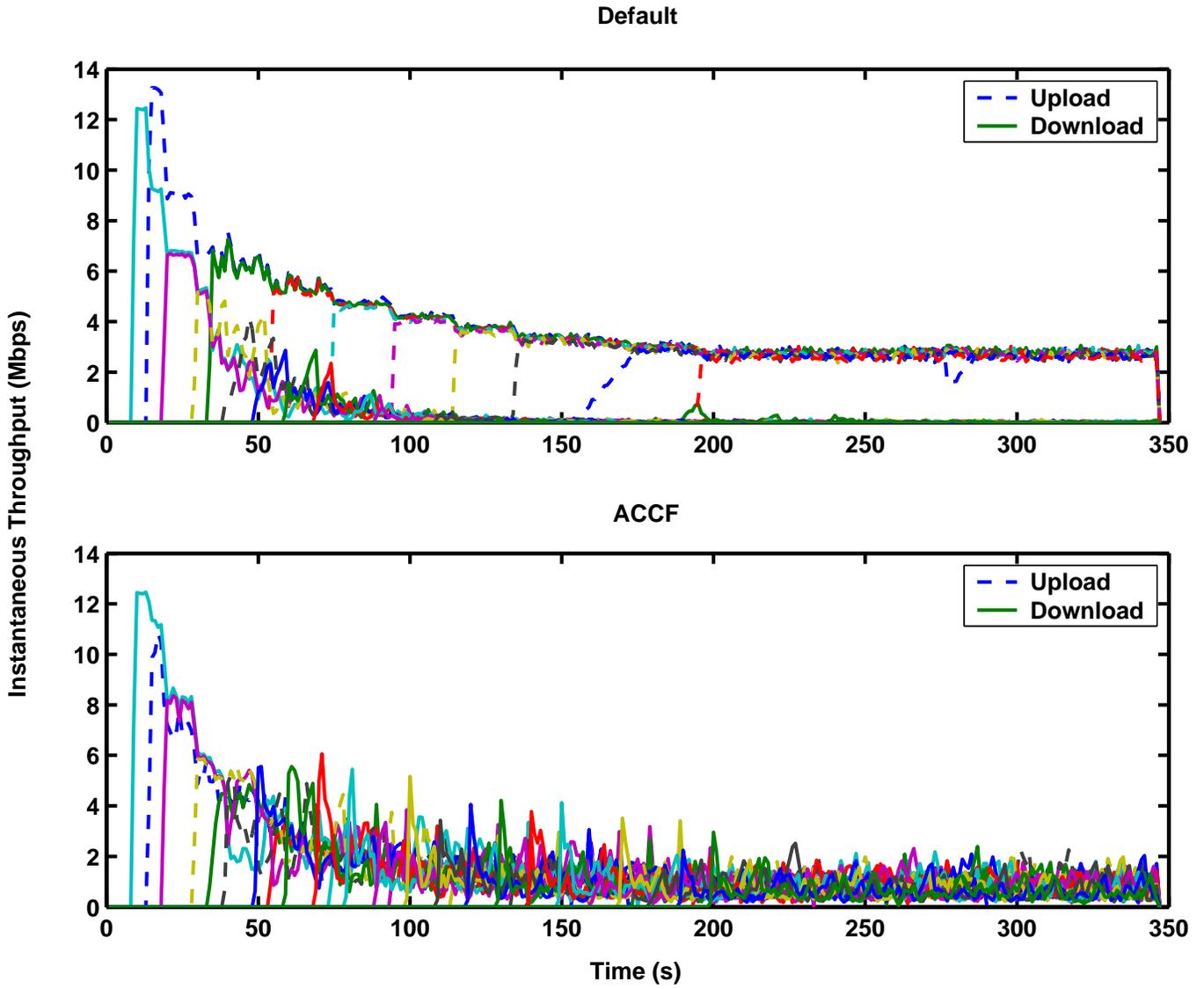}
\caption{Individual instantaneous throughput for upload and
download TCP flows when TCP receivers employ the delayed ACK
mechanism.} \label{fig:delayedack}
\end{figure}

We start download and upload FTP connections in 10 s and 20 s
intervals, respectively. Fig. \ref{fig:delayedack} shows the
instantaneous throughput for individual TCP flows over simulation
duration. As the results imply, in the default case, TCP download
connections starve in terms of throughput as the number of TCP
upload connections increase. In the meantime, some upload flows
experience long delays in starting and achieving high throughput
while some do not. On the other hand, using the proposed ACCF
scheme, all uplink and downlink TCP flows enjoy fair access. The
results are important in showing the proposed algorithm's
effectiveness even when the delayed TCP ACK mechanism is used.

\paragraph{Wireless Channel Errors}

In the third set of experiments, we assume the wireless channel to
be an Additive White Gaussian Noise (AWGN) channel. On top of the
energy-based PHY model of ns-2, we implemented a BER-based PHY
model according to the framework presented in \cite{Qijao01} using
the way of realization in \cite{Lacage06}. Our model considers the
channel noise power in Signal-to-Noise Ratio (SNR).
%A packet is
%detected at the receiver if the received power is over a specified
%threshold. Then, the packet error probability is calculated using
%the theoretical model presented in \cite{Qijao01} regarding the
%channel SNR, the modulation type used for the transmission of the
%packet (considering the different modulation of the headers and
%the payload), and the packet size.
%\footnotetext{If the energy of the first bit of the received
%packet is larger than a threshold, the reception is locked on the
%packet. If any other packet is received during the reception of
%another packet and the ratio of first bit energies are lower than
%a threshold, the packet is discarded. Otherwise, the collision
%routine is called.}
We set wireless channel noise levels such that each station
experience a finite data packet error rate (PER). We repeat the
tests for AWGN channel SNR values when PER is 0.001 or 0.01. We
only present the results on fairness index for the case when PER
is 0.01, since the results slightly differ and a similar
discussion holds for the case when PER is 0.001.

\begin{figure}[t]
\centering \includegraphics[width =
1.0\linewidth]{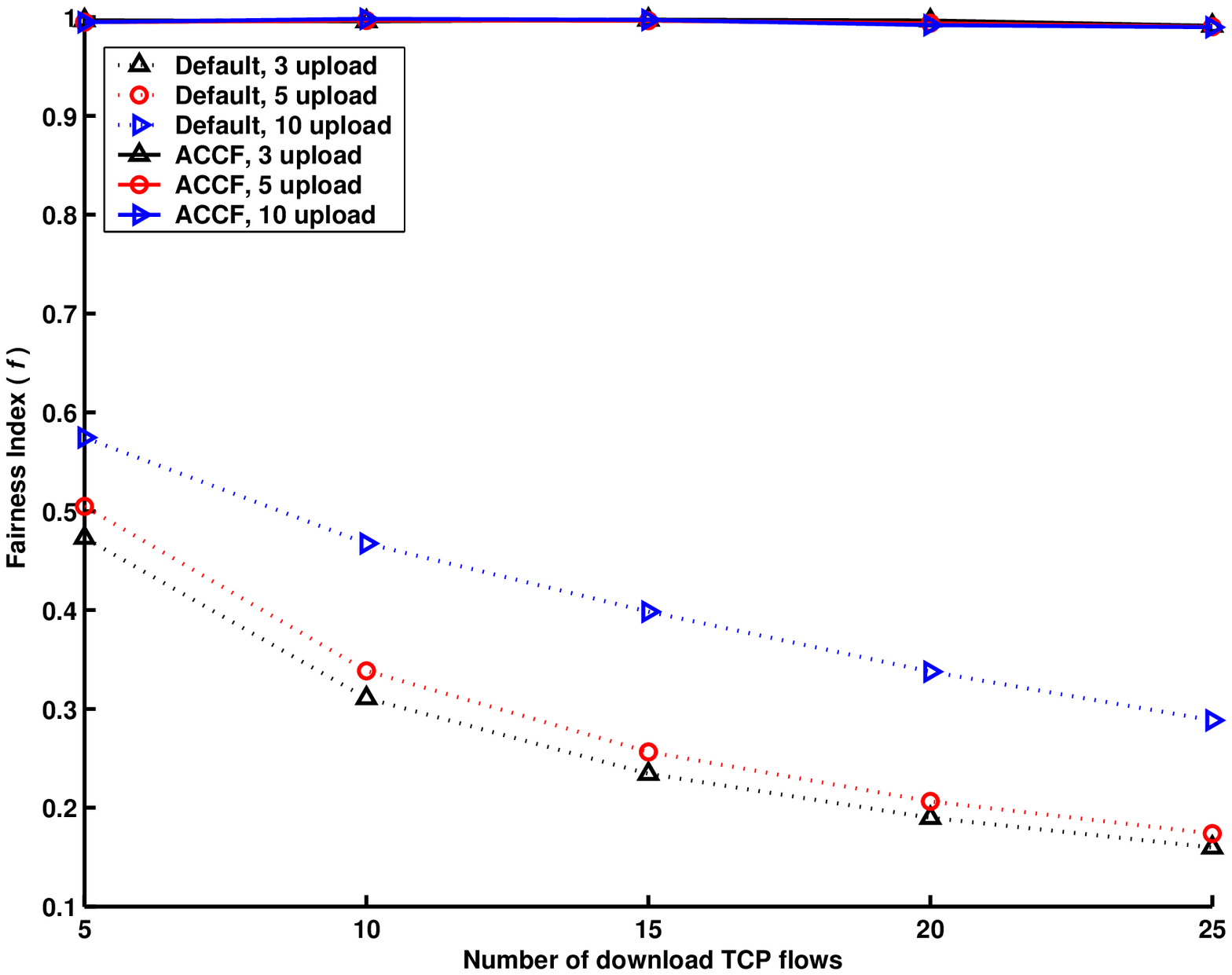} \caption{Fairness index among
all TCP flows over an AWGN channel with 0.1\% PER,when 3, 5, or 10
upload TCP connections are generated and the number of download
TCP connections are varied from 5 to 30.}
\label{fig:phyerrorpointone}
\end{figure}

\begin{figure}[t]
\centering \includegraphics[width =
1.0\linewidth]{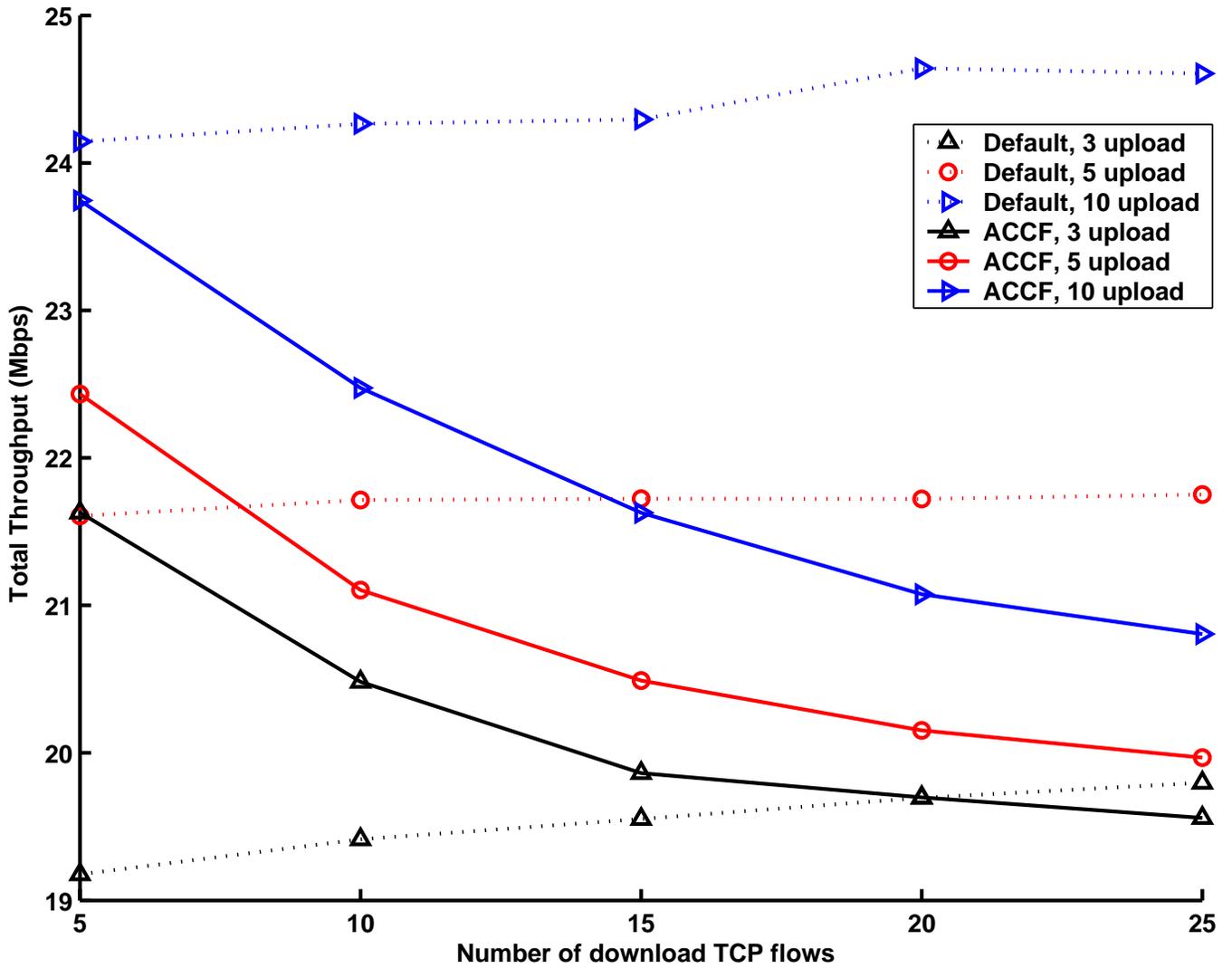} \caption{Total throughput of TCP
flows over an AWGN channel with 0.1\% PER, when 3, 5, or 10 upload
TCP connections are generated and the number of download TCP
connections are varied from 5 to 30.}
\label{fig:thput_phyerrorpointone}
\end{figure}

As in the first set of experiments, we generate 3, 5, or 10 upload
TCP connections and vary the number of download TCP connections
from 5 to 30. Fig. \ref{fig:phyerrorpointone} shows that the
proposed ACCF scheme provides fair access. The performance of ACCF
is resilient to wireless channel errors, i.e., fair access is
preserved even when there are errors in the wireless channel. As
shown in Fig.~\ref{fig:thput_phyerrorpointone}, the throughput
drops slightly when compared to errorless wireless channel case
due to the MAC retransmissions. Still, high channel utilization
maintained. DCF has slightly higher channel utilization at the
expense of fair access.

\paragraph{Varying source packet rates among TCP connections}

In the forth set of experiments, we test the performance when half
of the stations use the FTP agent, while the other half use the
Telnet agent with packet rates between 150 Kbps and 550 Kbps.

\begin{figure}[t]
\centering \includegraphics[width =
1.0\linewidth]{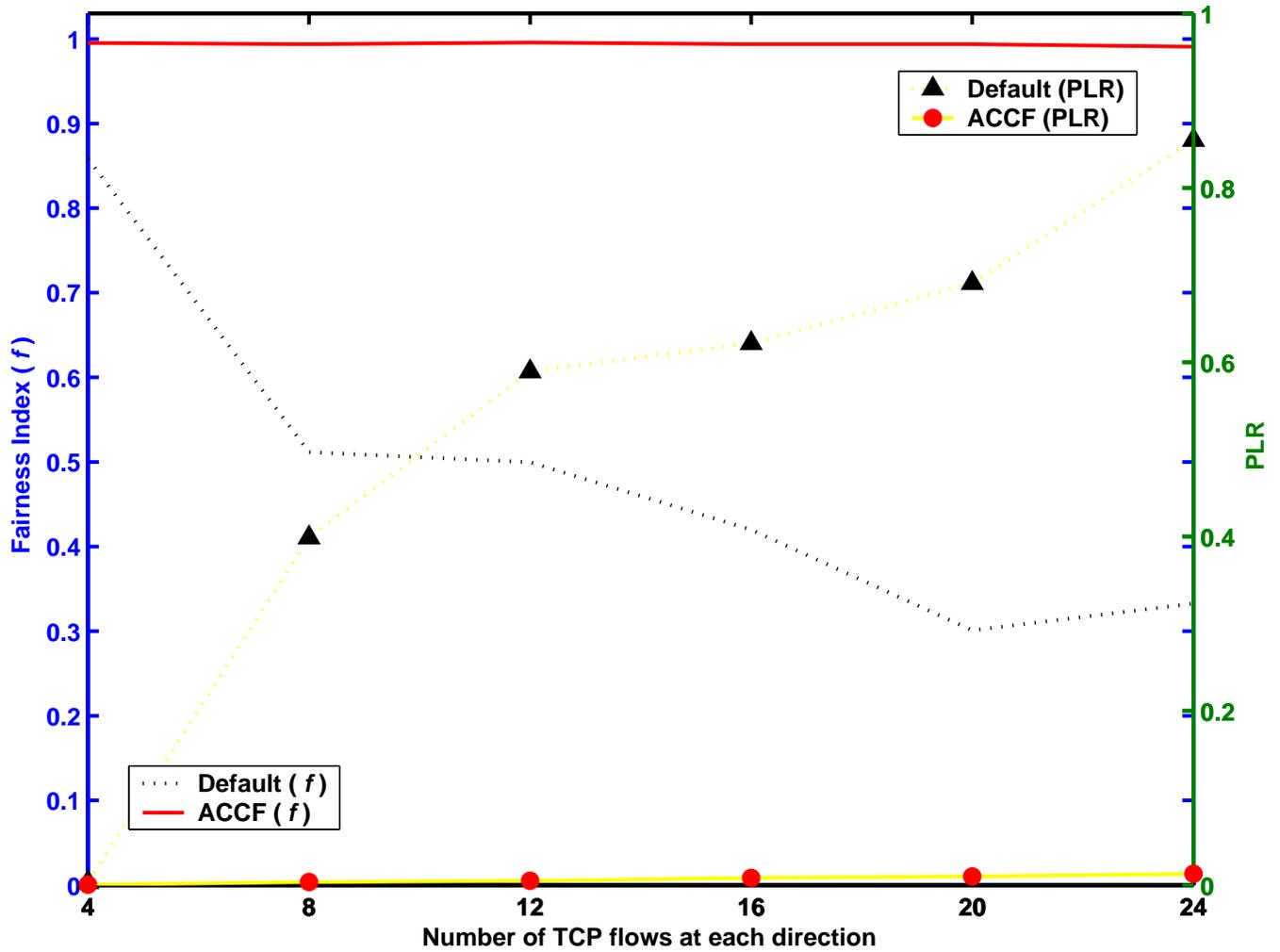} \caption{Fairness index
{\it f} for saturated and Packet Loss Rate (PLR) for unsaturated
TCP flows when the default EDCA or ACCF is employed.}
\label{fig:TCPfairness_nonsat_nodACK}
\end{figure}

\begin{figure}[t]
\centering \includegraphics[width =
1.0\linewidth]{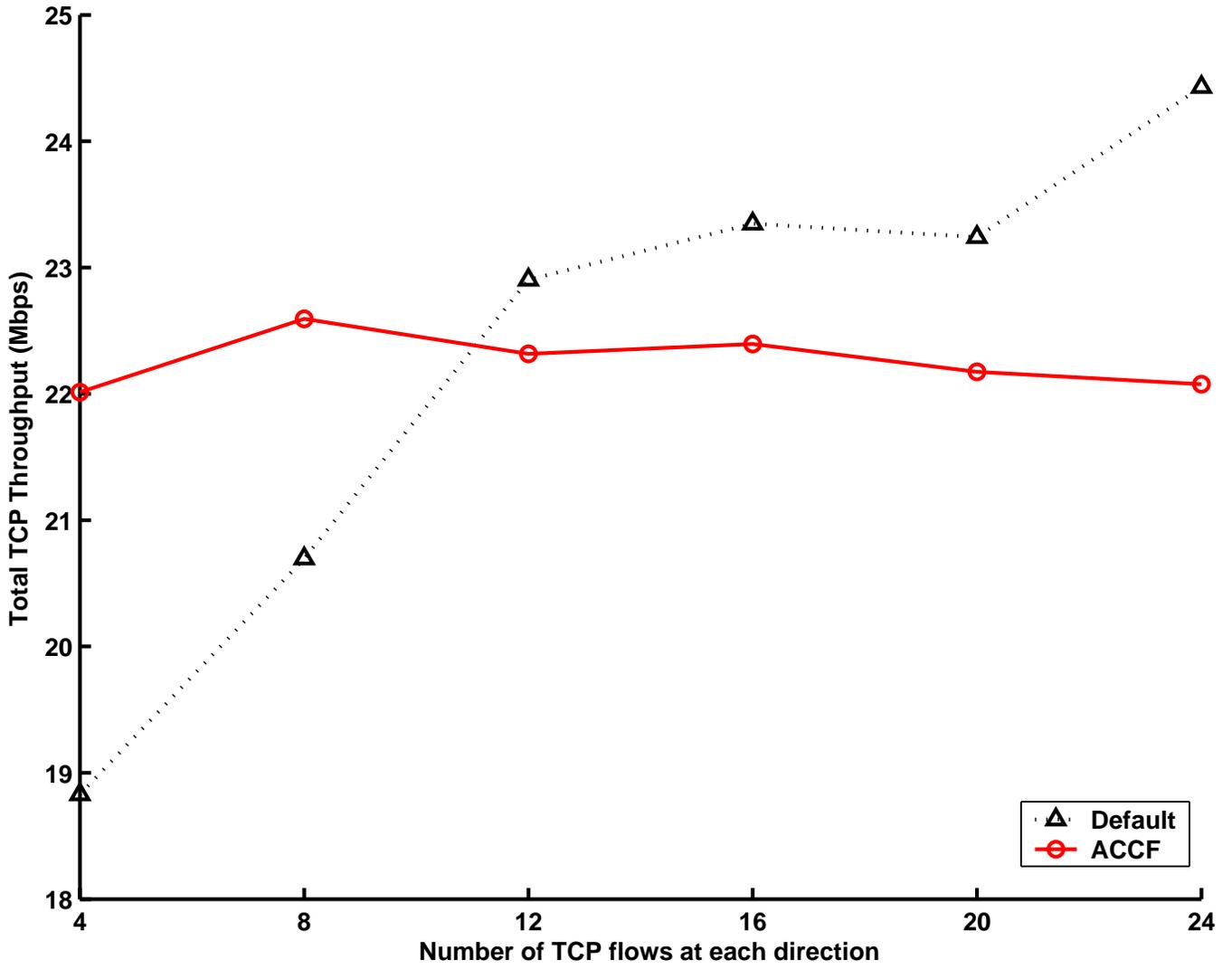} \caption{Total TCP
throughput when the default EDCA or ACCF is employed.}
\label{fig:TCPthroughput_nonsat_nodACK}
\end{figure}

Fig.~\ref{fig:TCPfairness_nonsat_nodACK} and
Fig.~\ref{fig:TCPthroughput_nonsat_nodACK} compares the
performance in terms of fair access and total throughput for
default DCF and ACCF for increasing number of TCP stations in each
direction, respectively. In,
Fig.~\ref{fig:TCPfairness_nonsat_nodACK}, the right y-axis denotes
the fairness index, $f$, among the FTP (saturated) flows, while
the left y-axis denotes the average Packet Loss Rate (PLR) for
Telnet (nonsaturated) flows. As the results present, the proposed
ACCF scheme can provide fair access (i.e., $f=1$ and $PLR=0$)
irrespective of the number of stations. As
Fig.~\ref{fig:TCPthroughput_nonsat_nodACK} shows, high channel
utilization is also maintained.

\begin{figure}[t]
\centering \includegraphics[width = 1.0\linewidth]{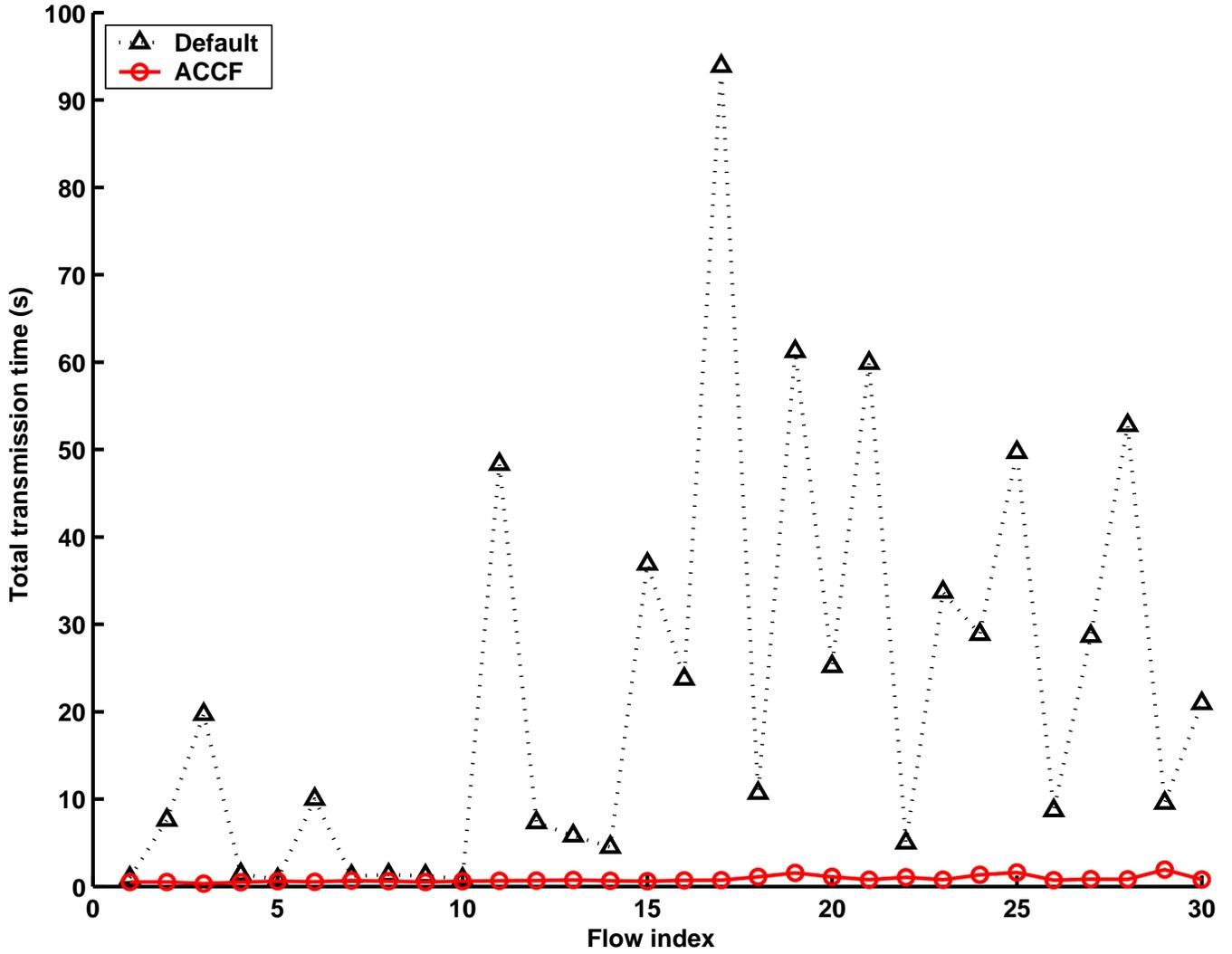}
\caption{The total transmission duration for individual
short-lived TCP flows.} \label{fig:shortterm}
\end{figure}

\paragraph{Short-lived TCP flows} In the fifth set of experiments,
we test the performance in terms of short-term fairness. First, we
generate 5 uplink and 10 downlink long-lived FTP flows. Then, 15
short-lived uplink and downlink FTP flows are generated with 5 s
intervals consecutively. Fig.\ref{fig:shortterm} shows the total
transmission duration for individual short-lived FTP flows for the
proposed ACCF algorithm and the default DCF. Note that the flow
indices from 1 to 15 represent uplink FTP flows while flow indices
from 16 to 30 represent downlink FTP flows. As the results imply,
the short-lived file transfer can be completed in a significantly
shorter time when the proposed algorithm is used. We can conclude
that the proposed ACCF algorithm is short-term fair. Although not
explicitly presented, most of the downlink connections experience
connection timeouts and even cannot complete the whole transaction
within the simulation duration for the default case.

\paragraph{Varying TCP Congestion Windows among Connections}
In the sixth set of experiments, we generate TCP connections with
receiver advertised congestion window sizes of 12, 20, 42, or 84.
We vary the number of FTP connections from 4 to 24 and the wired
link delay from 0 to 50 ms. For each scenario, the number of flows
using a specific congestion window size is uniformly distributed
among the connections (i.e., when there are 12 upload and 12
download TCP flows, 3 of the upload/download TCP connections use
the congestion window size \textit{W}, where \textit{W} is
selected from the set $S={12, 20, 42, 84}$). The wireless channel
is assumed to be errorless. The TCP delayed ACK mechanism is
enabled.

\begin{figure}[t]
\centering \includegraphics[width =
0.8\linewidth]{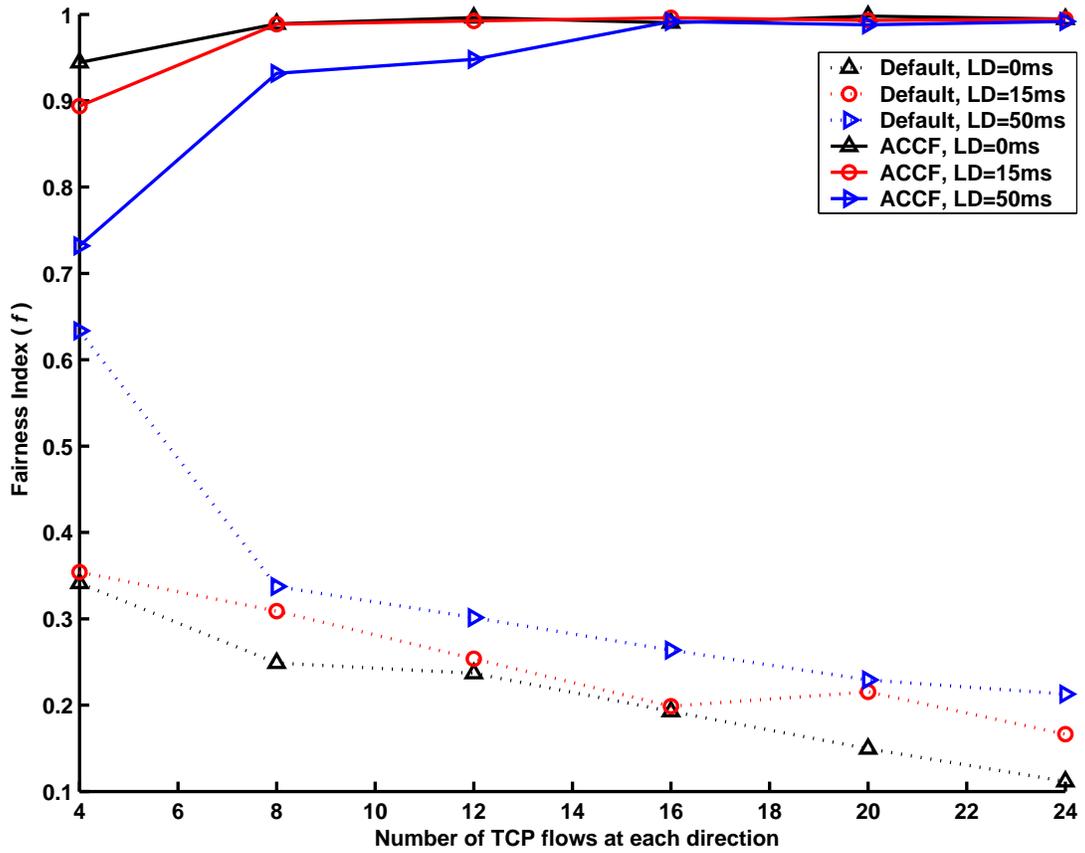} \caption{Fairness
index among all TCP flows with different congestion window sizes.}
\label{fig:TCPfairness_diffCWin1scenario}
\end{figure}

Fig. \ref{fig:TCPfairness_diffCWin1scenario} shows the fairness
index among all connections. We compare the default DCF results
with the results obtained when the AP employs the proposed ACCF.
As the results imply, with the introduction of the proposed
control block at the AP, a better fair resource allocation can be
achieved. However, a perfect fairness is not observed when the
link delay is larger and the number of flows is smaller. In these
cases, the bandwidth-delay product is larger than the receiver
advertised TCP congestion window size for connections with small
congestion windows. As a result, the throughput is limited by the
congestion window itself, not by the network bandwidth.

\begin{figure}
\centering \includegraphics[width =
0.8\linewidth]{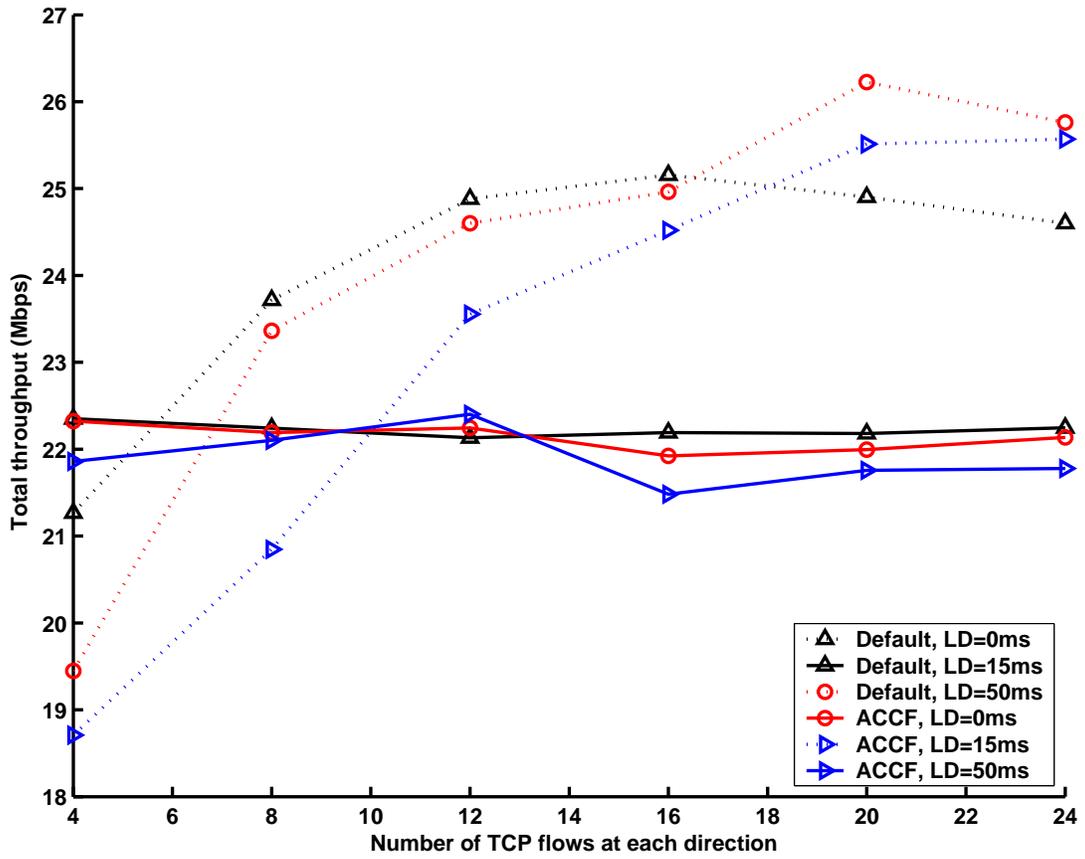} \caption{Throughput of
TCP connections when they use different congestion window sizes.}
\label{fig:TCPthput_diffCWin1scenario}
\end{figure}

In Fig. \ref{fig:TCPthput_diffCWin1scenario}, we plot the total
TCP throughput. As the results show, the default DCF has higher
channel utilization than ACCF. On the other hand, this comes at
the expense of fair access as shown in Fig.
\ref{fig:TCPfairness_diffCWin1scenario}. Although not explicitly
shown in Fig. \ref{fig:TCPthput_diffCWin1scenario}, all TCP
downlink flows are shut down when default DCF is employed (see
Section \ref{subsec:problem}). In this case, the shared channel
can mainly be utilized by data packets of uplink TCP connections.
In a fair scenario, as for ACCF, the TCP ACKs of downlink
connections sharing the channel are considerably higher in number
than the default DCF. As MAC efficiency decreases when packets of
shorter length access the channel, ACCF channel utilization
efficiency is slightly lower than DCF. The difference is more
notable for higher number of flows, since the AP, which is the
main source of short TCP ACK packet transmissions for the default
DCF, has a smaller share of the bandwidth.

\section{Conclusion}\label{sec:conclusion}

In this paper, we focused on unfair TCP access problem in an IEEE
802.11 infrastructure BSS. We have presented a novel and simple
analytical model to calculate the TCP congestion window limit that
provides fair TCP access in a wired/wireless scenario. The key
contribution of this study is that the proposed analytical model
considers varying wired link delays among connections, varying
number of uplink and downlink connections, and the use of delayed
ACK mechanism. Via simulations, we have shown that the congestion
window limits calculated via the proposed analysis (FCWA) provides
fair TCP access and high channel utilization. The same model can
also be used to decide on the required AP buffer size for fair TCP
access given the TCP congestion window limits used by the
connections. The cycle time analysis can be extended for IEEE
802.11e WLANs \cite{802.11e} as in \cite{Inan07_Globecom_Cycle},
therefore the analysis in this paper can also be extended for the
case when MAC parameter differentiation is used.

We have also designed a novel link layer access control block for
the AP that provides fair TCP access in an 802.11 infrastructure
BSS. Our simple idea for resolving the unfairness problem in the
WLAN is prioritizing TCP data packets of uplink flows over TCP ACK
packets of uplink flows at the AP. This idea originates from the
main finding of the proposed analytical model which shows that
fair access can be achieved by throttling TCP traffic (i.e.,
limiting congestion windows). The proposed link layer access
control block employs an ACK congestion control and filtering
(ACCF) algorithm. The proposed ACCF algorithm is unique in that
the specific algorithm parameters are based on the measured
average data transmission rate at the AP. Via simulations, we show
that fair resource allocation for uplink and downlink TCP flows
can be provided in a wide range of practical scenarios when the
proposed ACCF method is used. A key insight that can be obtained
from this study is that fair and efficient TCP access in a WLAN
can simply be achieved by intelligently scheduling TCP ACK
transmissions at the AP. As an attractive feature, ACCF does not
require any changes in the 802.11 standard, nor any enhancement at
the stations.

%GATHER{C:/INANCINAN/bibliography/standards.bib}
%GATHER{C:/INANCINAN/bibliography/HCCA.bib}
%GATHER{C:/INANCINAN/bibliography/simulations.bib}
%GATHER{C:/INANCINAN/bibliography/channel.bib}
%GATHER{C:/INANCINAN/bibliography/books.bib}
%GATHER{C:/INANCINAN/bibliography/EDCAanalysis.bib}
%GATHER{C:/INANCINAN/bibliography/mypapers.bib}
%GATHER{C:/INANCINAN/bibliography/fairness.bib}
%GATHER{C:/INANCINAN/bibliography/myreports.bib}
%GATHER{C:/INANCINAN/bibliography/TCPrelated.bib}
%GATHER{TCPfairness.bbl}

\bibliographystyle{IEEEtran}
\bibliography{IEEEabrv,C:/INANCINAN/bibliography/standards,C:/INANCINAN/bibliography/HCCA,C:/INANCINAN/bibliography/simulations,C:/INANCINAN/bibliography/channel,C:/INANCINAN/bibliography/books,C:/INANCINAN/bibliography/EDCAanalysis,C:/INANCINAN/bibliography/mypapers,C:/INANCINAN/bibliography/fairness,C:/INANCINAN/bibliography/myreports,C:/INANCINAN/bibliography/TCPrelated}

\begin{thebibliography}{10}
\providecommand{\url}[1]{#1}
\def\UrlFont{\rmfamily}
\providecommand{\newblock}{\relax}
\providecommand{\bibinfo}[2]{#2}
\providecommand\BIBentrySTDinterwordspacing{\spaceskip=0pt\relax}
\providecommand\BIBentryALTinterwordstretchfactor{4}
\providecommand\BIBentryALTinterwordspacing{\spaceskip=\fontdimen2\font plus
\BIBentryALTinterwordstretchfactor\fontdimen3\font minus
  \fontdimen4\font\relax}
\providecommand\BIBforeignlanguage[2]{{%
\expandafter\ifx\csname l@#1\endcsname\relax
\typeout{** WARNING: IEEEtran.bst: No hyphenation pattern has been}%
\typeout{** loaded for the language `#1'. Using the pattern for}%
\typeout{** the default language instead.}%
\else
\language=\csname l@#1\endcsname
\fi
#2}}

\bibitem{802.11}
\emph{{IEEE Standard 802.11: Wireless {LAN} medium access control (MAC) and
  physical layer (PHY) specifications}}, {IEEE 802.11} Std., 1999.

\bibitem{Balakrishnan99}
H.~Balakrishnan, V.~Padmanabhan, and R.~H. Katz, ``{The Effects of Asymmetry on
  TCP Performance},'' \emph{ACM Baltzer Mobile Networks and Applications
  (MONET)}, 1999.

\bibitem{Chiu89}
D.~Chiu and R.~Jain, ``{Analysis of the Increase/Decrease Algorithms for
  Congestion Avoidance in Computer Networks},'' \emph{Journal of Computer
  Networks and ISDN}, pp. 1--14, June 1989.

\bibitem{Vaidya00}
N.~H. Vaidya, P.~Bahl, and S.~Gupta, ``{Distributed Fair Scheduling in a
  Wireless LAN},'' in \emph{Proc. ACM Mobicom '00}, August 2000.

\bibitem{Nandagopal00}
T.~Nandagopal, T.~Kim, X.~Gao, and V.~Bharghavan, ``{Achieving MAC Layer
  Fairness in Wireless Packet Networks},'' in \emph{Proc. ACM Mobicom '00},
  August 2000.

\bibitem{802.11e}
\emph{{IEEE Standard 802.11: Wireless {LAN} medium access control (MAC) and
  physical layer (PHY) specifications: Medium access control (MAC) Quality of
  Service (QoS) Enhancements}}, {IEEE 802.11e} Std., 2005.

\bibitem{Casetti04}
C.~Casetti and C.~F. Chiasserini, ``{Improving Fairness and Throughput for
  Voice Traffic in 802.11e EDCA},'' in \emph{Proc. IEEE PIMRC '04}, September
  2004.

\bibitem{Leith05}
D.~J. Leith, P.~Clifford, D.~Malone, and A.~Ng, ``{TCP Fairness in 802.11e
  WLANs},'' \emph{{IEEE} Commun. Lett.}, pp. 964--966, November 2005.

\bibitem{Freitag06}
J.~Freitag, N.~L.~S. da~Fonseca, and J.~F. de~Rezende, ``{Tuning of 802.11e
  Network Parameters},'' \emph{{IEEE} Commun. Lett.}, pp. 611--613, August
  2006.

\bibitem{Tinnirello05_2}
I.~Tinnirello and S.~Choi, ``{Efficiency Analysis of Burst Transmissions with
  Block ACK in Contention-Based 802.11e WLANs},'' in \emph{Proc. IEEE ICC '05},
  May 2005.

\bibitem{Keceli08_ICCEDCA}
F.~Keceli, I.~Inan, and E.~Ayanoglu, ``{Weighted Fair Uplink/Downlink Access
  Provisioning in IEEE 802.11e WLANs},'' in \emph{IEEE ICC '08, Beijing,
  China}, May 2008.

\bibitem{SWKim05}
S.~W. Kim, B.-S. Kim, and Y.~Fang, ``{Downlink and Uplink Resource Allocation
  in IEEE 802.11 Wireless LANs},'' \emph{{IEEE} Trans. Veh. Technol.}, pp.
  320--327, January 2005.

\bibitem{Jeong05}
J.~Jeong, S.~Choi, and C.-K. Kim, ``{Achieving Weighted Fairness between Uplink
  and Downlink in IEEE 802.11 DCF-based WLANs},'' in \emph{Proc. IEEE QSHINE
  '05}, August 2005.

\bibitem{Pilosof03}
S.~Pilosof, R.~Ramjee, D.~Raz, Y.~Shavitt, and P.~Sinha, ``{Understanding TCP
  Fairness over Wireless LAN},'' in \emph{Proc. IEEE Infocom '03}, April 2003.

\bibitem{Wu05}
Y.~Wu, Z.~Niu, and J.~Zheng, ``{Study of the TCP Upstream/Downstream Unfairness
  Issue with Per-flow Queueing over Infrastructure-mode WLANs},''
  \emph{Wireless Commun. and Mobile Comp.}, pp. 459--471, June 2005.

\bibitem{Ha06}
J.~Ha and C.-H. Choi, ``{TCP Fairness for Uplink and Downlink Flows in
  WLANs},'' in \emph{Proc. IEEE Globecom '06}, November 2006.

\bibitem{Melazzi05}
N.~Blefari-Melazzi, A.~Detti, I.~Habib, A.~Ordine, and S.~Salsano, ``{TCP
  Fairness Issues in IEEE 802.11 Networks: Problem Analysis and Solutions Based
  on Rate Control},'' \emph{{IEEE} Trans. Wireless Commun.}, pp. 1346--1355,
  April 2007.

\bibitem{Keceli07_ICC}
F.~Keceli, I.~Inan, and E.~Ayanoglu, ``{TCP ACK Congestion Control and
  Filtering for Fairness Provision in the Uplink of IEEE 802.11 Infrastructure
  Basic Service Set},'' in \emph{Proc. IEEE ICC '07}, June 2007.

\bibitem{RFC3449}
``{RFC3449 - TCP Performance Implications of Network Path Asymmetry},'' 2002.

\bibitem{RFC2760}
``{RFC2760 - Ongoing TCP Research Related to Satellites},'' 2000.

\bibitem{Balakrishnan97}
H.~Balakrishnan, V.~N. Padmanabhan, and R.~H. Katz, ``{The Effects of Asymmetry
  on TCP Performance},'' in \emph{Proc. ACM/IEEE MobiCom '97}, November 1997.

\bibitem{Lakshman97}
T.~V. Lakshman, U.~Madhow, and B.~Suter, ``{Window-based Error Recovery and
  Flow Control with a Slow Acknowledgement Channel: A Study of TCP/IP
  Performance},'' in \emph{Proc. IEEE Infocom '97}, April 1997.

\bibitem{Fairhurst01}
G.~Fairhurst, N.~K.~G. Samaraweera, M.~Sooriyabandara, H.~Harun, K.~Hodson, and
  R.~Donadio, ``{Performance Issues in Asymmetric TCP Service Provision using
  Broadband Satellite},'' \emph{{IEE} Proc. Commun.}, pp. 95--99, 2001.

\bibitem{Allman98}
M.~Allman, ``{On the Generation and Use of TCP Acknowledgements},'' \emph{ACM
  Comp. Comm. Review}, October 1998.

\bibitem{Kalampouskas98}
L.~Kalampouskas, A.~Varma, and K.~K. Ramakrishnan, ``{Improving TCP Throughput
  over Two-Way Asymmetric Links: Analysis and Solutions},'' in \emph{Proc. ACM
  Sigmetrics '98}, 1998.

\bibitem{Zhang91}
L.~Zhang, S.~Shenker, and D.~D. Clark, ``{Observations on the Dyanmics of a
  Congestion Control Algorithm: The Effects of Two-Way Traffic},'' \emph{ACM
  Comp. Comm. Review}, pp. 133--147, 1991.

\bibitem{Medepalli05}
K.~Medepalli and F.~A. Tobagi, ``{Throughput Analysis of IEEE 802.11 Wireless
  LANs using an Average Cycle Time Approach},'' in \emph{Proc. IEEE Globecom
  '05}, November 2005.

\bibitem{Inan07_Globecom_Cycle}
I.~Inan, F.~Keceli, and E.~Ayanoglu, ``{Performance Analysis of the IEEE
  802.11e Enhanced Distributed Coordination Function using Cycle Time
  Approach},'' in \emph{Proc. IEEE Globecom '07}, November 2007.

\bibitem{ns2}
\BIBentryALTinterwordspacing
(2006) {The Network Simulator, ns-2}. [Online]. Available:
  \url{http://www.isi.edu/nsnam/ns}
\BIBentrySTDinterwordspacing

\bibitem{802.11g}
\emph{{IEEE Standard 802.11: Wireless {LAN} medium access control (MAC) and
  physical layer (PHY) specifications: Further Higher Data Rate Extension in
  the 2.4 GHz Band}}, {IEEE 802.11g} Std., 2003.

\bibitem{Jain91}
R.~Jain, \emph{{The Art of Computer Systems Performance Analysis: Techniques
  for Experimental Design, Measurement, Simulation, and Modeling}}.\hskip 1em
  plus 0.5em minus 0.4em\relax John Wiley and Sons, 1991.

\bibitem{Keceli08_fairTCP_trep}
F.~Keceli, I.~Inan, and E.~Ayanoglu, ``{Fair TCP Access Provisioning in the
  IEEE 802.11 Infrastructure Basic Service Set},'' Tech. Rep., May 2008.

\bibitem{Qijao01}
D.~Qijao and S.~Choi, ``{Goodput Enhancement of IEEE 802.11a Wireless LAN via
  Link Adaptation},'' in \emph{Proc. IEEE ICC '01}, June 2001.

\bibitem{Lacage06}
\BIBentryALTinterwordspacing
M.~Lacage. (2006) {Ns-2 802.11 Support}. INRIA Sophia Antipolis. France.
  [Online]. Available: \url{http://spoutnik.inria.fr/code/ns-2}
\BIBentrySTDinterwordspacing

\end{thebibliography}

\clearpage

\clearpage

%\clearpage
%
%\begin{figure}[t]
%\centering \includegraphics[width =
%1.0\linewidth]{TCPfairness_onlydownload} \caption{Fairness index
%among all TCP download flows when the number of download TCP
%connections is varied from 5 to 30.}
%\label{fig:TCPfairness_onlydownload}
%\end{figure}

\clearpage

\clearpage

%\clearpage
%
%\begin{figure}[t]
%\centering \includegraphics[width =
%1.0\linewidth]{TCPthput_15flows_equalLD_onlydownload}
%\caption{Individual throughput of TCP download flows.}
%\label{fig:thput_15downflows}
%\end{figure}

\clearpage

\clearpage

\clearpage

\clearpage

\clearpage

\clearpage

\clearpage

\clearpage

\clearpage

\clearpage

\clearpage

%\begin{figure}[t]
%\centering \includegraphics[width =
%1.0\linewidth]{TCPfairness_per00} \caption{Fairness index among
%all TCP flows over an errorless channel, when 3, 5, or 10 upload
%TCP connections are generated and the number of download TCP
%connections are varied from 5 to 30.} \label{fig:phyerrorzero}
%\end{figure}
%
%\clearpage
%
%\begin{figure}[t]
%\centering \includegraphics[width = 1.0\linewidth]{TCPthput_per00}
%\caption{Total throughput of TCP flows over an errorless channel,
%when 3, 5, or 10 upload TCP connections are generated and the
%number of download TCP connections are varied from 5 to 30.}
%\label{fig:thput_phyerrorzero}
%\end{figure}
%
%\clearpage
%
%\begin{figure}[t]
%\centering \includegraphics[width =
%1.0\linewidth]{TCPfairness_per001} \caption{Fairness index among
%all TCP flows over an AWGN channel with 0.01\% PER, when 3, 5, or
%10 upload TCP connections are generated and the number of download
%TCP connections are varied from 5 to 30.}
%\label{fig:phyerrorpointzeroone}
%\end{figure}
%
%\clearpage
%
%\begin{figure}[t]
%\centering \includegraphics[width =
%1.0\linewidth]{TCPthput_per001} \caption{Total throughput of TCP
%flows over an AWGN channel with \%0.01 PER, when 3, 5, or 10
%upload TCP connections are generated and the number of download
%TCP connections are varied from 5 to 30.}
%\label{fig:thput_phyerrorpointzeroone}
%\end{figure}

\clearpage

\clearpage

\clearpage

\clearpage

%\clearpage
%
%\begin{figure}[t]
%\centering \includegraphics[width =
%1.0\linewidth]{TCPfairness_nonsat_dACK} \caption{Fairness index
%{\it f} for saturated and Packet Loss Rate (PLR) for unsaturated
%TCP flows when the default EDCA or ACCF is employed and delayed
%ACK mechanism is used ($b=2$).}
%\label{fig:TCPfairness_nonsat_dACK}
%\end{figure}
%
%\clearpage
%
%\begin{figure}[t]
%\centering \includegraphics[width =
%1.0\linewidth]{TCPthroughput_nonsat_dACK} \caption{Total TCP
%throughput when the default EDCA or ACCF is employed and delayed
%ACK mechanism is used ($b=2$).}
%\label{fig:TCPthroughput_nonsat_dACK}
%\end{figure}

\clearpage

\clearpage

\clearpage

\end{document}